\documentclass[useAMS, usenatbib, usegraphicx]{mn2e}
\usepackage{amssymb}
\usepackage[colorlinks=TRUE, dvips]{hyperref}
\def\rpd{\hbox{rad\,d$^{-1}$}}
\def\mrpd{\hbox{mrad\,d$^{-1}$}}
\def\chisq{\hbox{$\chi^2$}}
\def\chisqr{\hbox{$\chi^2_r$}}
\def\msun{\hbox{${\rm M}_{\odot}$}}
\def\rsun{\hbox{${\rm R}_{\odot}$}}
\def\mstar{\hbox{$M_{\star}$}}
\def\rstar{\hbox{$R_{\star}$}}
\def\teff{\hbox{$T_{\rm eff}$}}
\def\sn{\hbox{S/N}}

\def\ms{\hbox{m\,s$^{-1}$}}
\def\kms{\hbox{km\,s$^{-1}$}}
\def\eps{\hbox{erg\,s$^{-1}$}}
\def\vsini{\hbox{$v\sin i$}}
\def\rsini{\hbox{$R\sin i$}}
\def\ptt{\hbox{$10^{-4} I_{\rm c}$}}
\def\degr{\hbox{$^\circ$}}
\def\Prot{\hbox{$P_{\rm rot}$}}
\def\mj{\hbox{$\rm M_J$}}
\def\d{\hbox{$\rm d$}}
\def\kG{\hbox{$\rm kG$}}
\def\Omeq{\hbox{$\Omega_{\rm eq}$}}
\def\dOm{\hbox{$d\Omega$}}
\def\IUR{\hbox{$I_{\rm UR}$}}
\def\Iq{\hbox{$I_{\rm q}$}}
\def\VUR{\hbox{$V_{\rm UR}$}}

\begin{document}

\title[Magnetic topologies of mid-M dwarfs] 
{Large-scale magnetic topologies of mid-M dwarfs\thanks{Based on
observations
obtained at the Canada-France-Hawaii  Telescope (CFHT) and at the
T\'elescope Bernard Lyot (TBL). CFHT is operated by the National
Research Council of Canada, the Institut National des Science de
l'Univers of the Centre National de la Recherche Scientifique of France
(INSU/CNRS), and the University of Hawaii, while the TBL is operated by
CNRS/INSU.} }

\def\newauthor{%
  \end{author@tabular}\par
  \begin{author@tabular}[t]{@{}l@{}}}
\makeatother
 
\author[J. Morin et al.] {\vspace{1.7mm}
J.~Morin$^1$\thanks{E-mail: 
jmorin@ast.obs-mip.fr (JM);
donati@ast.obs-mip.fr (J-FD);
petit@ast.obs-mip.fr (PP);
xavier.delfosse@obs.ujf-grenoble.fr (XD);
thierry.forveille@obs.ujf-grenoble.fr (TF);
albert@cfht.hawaii.edu (LA);
auriere@ast.obs-mip.fr (MA);
remi.cabanac@ast.obs-mip.fr (RC);
dintrans@ast.obs-mip.fr (BD);
rfares@ast.obs-mip.fr (RF);
tgastine@ast.obs-mip.fr (TG);
mmj@st-andrews.ac.uk (MMJ);
ligniere@ast.obs-mip.fr (FL);
fpaletou@ast.obs-mip.fr (FP);
jramirez@mesiog.obspm.fr (JR);
sylvie.theado@ast.obs-mip.fr (ST)},
J.-F.~Donati$^1$, P.~Petit$^1$, X.~Delfosse$^2$, T.~Forveille$^2$,
L.~Albert$^3$, \\ 
\vspace{1.7mm}
{\hspace{-1.5mm}\LARGE\rm 
M.~Auri\`ere$^1$, R.~Cabanac$^1$, B.~Dintrans$^1$, R.~Fares$^1$,
T.~Gastine$^1$, M.M.~Jardine$^4$,} \\
\vspace{1.7mm}
{\hspace{-1.5mm}\LARGE\rm 
F.~Ligni\`eres$^1$, F.~Paletou$^1$, J.C.~Ramirez Velez$^5$, S.~Th\'eado$^1$} \\
$^1$ LATT, Universit\'e de Toulouse, CNRS, 14 Av.\ E.~Belin, F--31400
Toulouse, France\\
$^2$ LAOG--UMR~5571, CNRS et Univ.\ J.~Fourier, 31 rue de la Piscine,
F--38041 Grenoble, France\\
$^3$ CFHT, 65-1238 Mamalahoa Hwy, Kamuela HI, 96743 USA\\
$^4$ School of Physics and Astronomy, Univ.\ of St~Andrews, St~Andrews,
Scotland KY16 9SS, UK \\
$^5$ LESIA, Observatoire de Paris-Meudon, 92195 Meudon, France}
\date{\today,~ $Revision: 1.64 $}
\maketitle
 
\begin{abstract}
We present in this paper the first results of a spectropolarimetric
analysis of a small sample ($\sim20$) of active stars ranging from
spectral type M0 to M8, which are either fully-convective or possess a
very small radiative core. This study aims at providing new constraints
on dynamo processes in fully-convective stars. 

The present paper focuses on 5 stars of spectral type $\sim$M4, i.e.
with masses close to the full convection threshold ($\simeq
0.35~\msun$), and with short rotational periods. Tomographic imaging
techniques allow us to reconstruct the surface magnetic topologies from
the rotationally modulated time-series of circularly polarised profiles.
We find that all stars host mainly axisymmetric large-scale poloidal
fields. Three stars were observed at two different epochs separated by
$\sim$1~yr; we find the magnetic topologies to be globally stable on
this timescale. 

We also provide an accurate estimation of the rotational period of all
stars, thus allowing us to start studying how rotation impacts the
large-scale magnetic field.

\end{abstract}

\begin{keywords} 
stars: magnetic fields --  
stars: low-mass -- 
stars: rotation -- 
techniques: spectropolarimetry 
\end{keywords}

\section{Introduction} 
\label{sec:intro}
Magnetic fields play a key role in every phase of the life of stars and
are linked to most of their manifestations of activity. Since
\cite{Larmor19} first proposed that electromagnetic induction might be
the origin of the Sun's magnetic field, dynamo generation of magnetic
fields in the Sun and other cool stars has been a subject of constant
interest. The paradigm of the $\alpha\Omega$ dynamo, i.e. the generation
of a large-scale magnetic field through the combined action of
differential rotation ($\Omega$ effect) and cyclonic convection
($\alpha$ effect), was first proposed by \cite{Parker55} and then
thoroughly debated and improved \citep[e.g.,][]{Babcock61, Leighton69}.
A decade ago, helioseismology provided the first measurements of the
internal differential rotation in the Sun and thus revealed a thin zone
of strong shear at the interface between the radiative core and the
convective envelope. During the past few years, theoreticians pointed
out the crucial role for dynamo processes of this interface -- called
the tachocline -- being the place where the $\Omega$ effect can amplify
magnetic fields (see \citealt{Charbonneau05} for a review of solar
dynamo models).

Among cool stars, those with masses lower than about 0.35~\msun\ are
fully-convective \citep[e.g.,][]{Chabrier97}, and therefore do not
possess a tachocline; some observations further suggest that they
rotate almost as rigid bodies \citep{Barnes05}. However, many
fully-convective stars are known to show various signs of activity such
as radio, Balmer line, and X-ray emissions \citep[e.g.,][]{Joy49,
Lovell63, Delfosse98, Mohanty03, West04}. Magnetic fields have been
directly detected thanks to Zeeman effect on spectral lines, either in
unpolarised light \citep[e.g.,][]{Saar85, Johns96, Reiners06}, or
in circularly polarised spectra \citep[][]{Donati06}.

The lack of a tachocline in very-low-mass stars led theoreticians to
propose non-solar dynamo mechanism in which cyclonic convection and
turbulence play the main roles while differential rotation only has
minor effects \citep[e.g.,][]{Durney93}. During past few years, several
semi-analytical approaches and MHD simulations were developed in order
to model the generation of magnetic fields in fully-convective
stars. Although they all conclude that fully-convective stars should be
able to produce a large-scale magnetic field, they disagree on the
properties of such a field, and the precise mechanisms involved in the
dynamo effect remain unclear. Mean-field modellings by \cite{Kuker05}
and \cite{Chabrier06} assumed solid-body rotation and found $\alpha^2$
dynamo generating purely non-axisymmetric large-scale fields. Subsequent
direct numerical simulations diagnose either ``antisolar''  differential
rotation (i.e. poles faster than the equator) associated
with a net axisymmetric poloidal field \citep[e.g.,][]{Dobler06}; or
strongly quenched ``solar'' differential rotation (i.e. the equator
faster than the poles) and a strong axisymmetric toroidal field
component \citep[e.g.,][]{Browning08}.

The first detailed observations of fully-convective stars do not
completely agree with any of these models. Among low-mass stars, differential
rotation appears to vanish with increasing convective depth
\citep{Barnes05}. This result is further confirmed by the first
detailed spectropolarimetric observations of the very active
fully-convective star V374~Peg by \cite{Donati06} and \cite{Morin08}
(hereafter M08) who measure very weak differential rotation (about
1/10th of the solar surface shear). These studies also report a strong
mostly axisymmetric poloidal surface magnetic field stable on a
timescale of 1~yr on V374~Peg, a result which does not completely
agree with any of the existing theoretical predictions. V374~Peg being a
very fast rotator, observations of fully-convective stars with longer
rotation periods are necessary to generalise these results.

In order to provide theoretical models and numerical simulations with
better constraints, it is necessary to determine the main magnetic
field properties -- topology and time-variability -- of several
fully-convective stars, and to find out their dependency on stellar
parameters -- mass, rotation rate, and differential rotation. In this
paper, we present and analyse the spectropolarimetric observations
of a small sample of stars just around the limit to full convection
(spectral types ranging from M3 to M4.5), collected with ESPaDOnS and 
NARVAL between 2006 Jan and 2008 Feb. Firstly, we briefly
present our stellar sample, and our observations are described in a
second part. We then provide insight on the imaging process and
associated physical model. Afterwards, we present our analysis for each
star of the sample. Finally, we discuss the global trends found in our
sample and their implications in the understanding of dynamo processes
in fully-convective stars. 

\section{Stellar sample}
\label{sec:sample}
Our stellar sample includes 5 M-dwarfs just about the full-convection
threshold i.e. around spectral type M4. It is part of a wider sample of
about 20 stars ranging from M0 to M8; results for remaining stars will 
be presented in forthcoming papers. The stars were selected from the
rotation-activity study of \cite{Delfosse98}. We chose only active stars
so that the magnetic field is strong enough to produce detectable
circularly polarised  signatures, allowing us to apply tomographic
imaging techniques. Stars with spin periods ranging from 0.4 to 4.4~\d\
were selected to study the impact of rotation on the reconstructed magnetic
topologies (though all the observed stars lie in the saturated regime, see
Sec.~\ref{sec:disc}).

\begin{table*}
\caption[]{Fundamental parameters of the stellar sample. Columns
1--8 respectively list the name, the spectral type \citep[taken
from][]{Reid95}, the stellar mass (see Sec.~\ref{sec:sample}), the
bolometric luminosity and log$R_X=$log($L_X/L_{bol}$) (see text), the
projected rotation velocity as inferred from Zeeman Doppler Imaging
(ZDI), and the rotation periods $P_{ZDI}$ (used to compute the
ephemeris) and \Prot\ (accurate period derived from our study). Columns
9--13 respectively list the empirical convective turnover time (see
text), the effective Rossby number (see text), the \rsini\ , the
theoretical radius suited to the stellar mass (see text), and the
inclination angle used for ZDI deduced by comparing columns 11 and 12.
For columns 8 and 11 we also mention, between brackets, respectively
3-$\sigma$ and 1-$\sigma$ error bars inferred from our study. For
the precision of the other quantities refer to Section~\ref{sec:sample}.}
 \begin{tabular}{ccccccccccccccc}
 \hline
 Name & ST & \mstar & log$L_{bol}$ & log$R_X$ & \vsini & $P_{\rm ZDI}$
& \Prot & $\tau_c$ & $Ro$ & \rsini & \rstar & $i$ \\
 & & (\msun) & (\eps) & & (\kms) & (d) & (d) & (d) &
($10^{-2}$) & (\rsun) & (\rsun) & (\degr) \\ 
 \hline
 AD Leo & M3 & $0.42$ & 31.91 & -3.18 & $3.0$ & 2.22 & $2.2399(6)$
& 48 & 4.7 & $0.13(4)$ & 0.38 & 20 \\
 EQ Peg A & M3.5 & $0.39$ & 31.84 & -3.02 & $17.5$ & $1.06$ &
$1.061(4)$ & 54 & 2.0 & $0.37(2)$ & 0.35 & 60 \\
 EV Lac & M3.5 & $0.32$ & 31.66 & -3.33 & $4.0$ & 4.378 &
$4.3715(6)$ & 64 & 6.8 & $0.35(9)$ & 0.30 & 60 \\
 YZ CMi & M4.5 & $0.31$ & 31.64 & -3.09 & $5.0$ & 2.77 &
$2.7758(6)$ & 66 & 4.2 & $0.27(5)$ & 0.29 & 60 \\
 V374~Peg & M4 & $0.28$ & 31.56 & -3.20 & $36.5$ & -- &
$0.445654(2)$ & 72 & 0.6 & $0.32(1)$ & 0.28 & 70 \\
 EQ Peg B & M4.5 & $0.25$ & 31.47 & -3.25 & $28.5$ & 0.405 & $0.404(4)$
& 76 & 0.5 & $0.23(1)$ & 0.25 & 60 \\
 \hline
 \end{tabular}
 \label{tab:sample}
\end{table*}

The analysis carried out in the present paper concerns: AD~Leo
(GJ~388) which is partly-convective, EV~Lac (GJ~873), YZ~ CMi (GJ~285),
EQ~Peg~A (GJ~896~A) which lies just on the theoretical limit to
full-convection, and EQ~Peg~B (GJ~896~B). All are known as active
flare-stars, and strong magnetic fields have already been reported for
some stars \citep[e.g.,][]{Saar85, Johns96, Reiners07}. We include the
previously studied M4 star V374~Peg in our analysis (M08).

The main properties of the stellar sample, inferred from the present
work or collected from previous ones, are shown in
Table~\ref{tab:sample}. We show stellar masses computed using the
empirical relation derived by \cite{Delfosse00} and based on J-band
absolute magnitude values inferred from apparent magnitude measurements
of 2MASS \citep[][]{Cutri03} and Hipparcos parallaxes \citep[][]{ESA97}.
For EQ~Peg~A and B, the values we find are in good agreement with the
dynamical mass of the binary system of $0.61\pm0.03$ reported by
\cite{Tamazian06}. Radius and bolometric luminosity suited to the
stellar mass are computed from NextGen models \citep[][]{Baraffe98}. We
also mention log$R_X=$ log($L_X/L_{bol}$), where $L_X$ is an average of
NEXXUS values (excluding outliers supposedly corresponding to flares).
We observe dispersions ranging from 0.1 to 0.2 in log($L_X$),
corresponding to intrinsic variability. As no data is available on
NEXXUS for EQ~Peg~B alone, we take one fourth of EQ~Peg~A's X-ray
luminosity, as reported by \cite{Robrade04}. Line of sight projected
equatorial velocities (\vsini), rotation periods (\Prot) and inclination
($i$) of the rotation axis with respect to the line of sight are
inferred from the present study. We estimate that the absolute accuracy
to which \vsini\ is determined is about 1~\kms. The uncertainty on
\Prot\ is precisely computed (see \ref{sec:mod-diffrot}). The
inclination angle estimate is coarse (accuracy of about 20\degr),
tomographic imaging does not require more precision. 

To study how activity and magnetic fields vary among stars of different 
masses, the most relevant parameter to consider is the effective Rossby 
number $Ro=\Prot/\tau_c$ (where $\tau_c$ is the convective turnover time, 
\citealt[e.g.,][]{Noyes84}).  We take convective turnover times from 
\citet[][empirically derived from X ray fluxes of M dwarfs]{Kiraga07};  
$\tau_c$ is found to increase strongly (as expected) with 
decreasing mass and bolometric luminosities.  
For the present sample, we find that $Ro$ ranges from 0.005 to 0.07, 
i.e., much smaller than in the Sun (where $Ro\simeq1.5-2.0$) as a 
result of both the shorter \Prot\ and the larger $\tau_c$ (see 
Table~\ref{tab:sample}).  

%% {\bf We note that small effective Rossby numbers ($Ro\sim 10^{-2}$)
%% are obtained in this M-dwarfs sample compared to, say, the Sun
%% where $Ro$ is rather of $\od{1}$ and it results from following both
%% points: (i) these M-stars rotate twenty five times as fast than the
%% Sun; (ii) their intrinsic luminosities are much smaller than the
%% solar one and hence also their typical convective velocities leading
%% to larger convective overturning times $\tau_c$. As we will show
%% in the discussion, the weakness of Rossby's number in these fully
%% convective M-stars is a key ingredient as regards the magnetic field
%% topology or the differential rotation.}

\section{Observations}
\label{sec:obs}
Spectropolarimetric observation of our 5 mid-M stars were collected
between 2006 Jan and 2008 Feb with the twin instruments
ESPaDOnS on the 3.6m Canada-France-Hawaii Telescope (CFHT) located in
Hawaii, and NARVAL on the 2m T\'elescope Bernard Lyot (TBL) in southern
France. ESPaDOnS and NARVAL are built from the same design
\citep{Donati03c}. They produce spectra spanning the entire optical
domain (from 370 to 1000~nm) at a resolving power of about 65~000. Each
observation consists of four individual subexposures taken in different
polarimeter configurations which are combined together so that all
spurious polarisation signatures are cancelled to first order
\citep[e.g.,][]{Donati97}.

Data reduction was carried out using \textsc{libre-esprit}. This fully
automated package/pipeline installed at CFHT and TBL performs optimal
extraction of NARVAL and ESPaDOnS unpolarised (Stokes $I$) and
circularly polarised (Stokes $V$) spectra, following the procedure
described in \cite{Donati97}. The peak signal-to-noise ratios (\sn ) per
2.6~\kms\ velocity bin range from 100 to 500, depending on the magnitude
of the target, the telescope used and the weather conditions. The full
journal of observations is presented in Tables~\ref{tab:log_adleo} to
\ref{tab:log_eqpegb}.

All spectra are automatically corrected for spectral shifts resulting
from instrumental effects (e.g., mechanical flexures, temperature or
pressure variations) using telluric lines as a reference. Though not
perfect, this procedure allows spectra to be secured with a radial
velocity (RV) precision of better than $0.030~\kms$
\citep[e.g.,][]{Moutou07}.

Least-squares deconvolution \citep[LSD, ][]{Donati97} was applied to all
observations, in order to extract the polarimetric information from most
photospheric atomic lines and gather it into a unique synthetic profile
of central wavelength $\lambda_0 = 700~{\rm nm}$ and effective Land\'e
factor $g_{\rm eff} = 1.2$. The line list for LSD was computed from an
Atlas9 local thermodynamic equilibrium model \citep{Kurucz93} matching
the properties properties of our whole sample, and contains about 5~000
moderate to strong atomic lines. We notice a multiplex gain of about 10
with respect to the \sn\ of the individual spectra of our sample. Zeeman
signatures are clearly detected in all the spectra (see
Sec.~\ref{sec:adleo} to \ref{sec:eqpegb})
with maximum amplitudes varying from 0.5\% (for EQ~Peg~B) to 1.2\% (for
AD~Leo) of the unpolarised continuum level. Temporal variations, due to
rotational modulation, of the Zeeman signatures is obvious for some
stars, whereas it is very weak on others, mostly depending on the
inclination angle of their rotation axis with respect to the line of
sight.

For each observation we compute the corresponding longitudinal magnetic
field (i.e. the line of sight projection) from the Stokes $I$ and $V$
LSD profiles through the relation :

\begin{equation}
 B_l({\rm G}) = -2.14 \times 10^{11} \frac{\displaystyle\int v\,V(v)
\,{\rm d}v}{\lambda_0\,g_{\rm eff}\,c \displaystyle\int
\left[I_c-I(v)\right] {\rm d}v } \, ,
\label{eq:bl}
\end{equation}
\citep[][]{Rees79, Donati97, Wade00} where $v$ is the radial velocity in
the star's rest frame, $\lambda_0$, in nm, is the mean wavelength of the
LSD profile, $c$ is the velocity of light in vacuum in the same unit as
$v$, $g_{\rm eff}$ is the value of the mean Land\'e factor of the
LSD line, and $I_c$ the continuum level.

In the rest of the paper, all data are phased according to the following
ephemeris:
\begin{equation}
\mbox{HJD} = 2453950.0 + P_{ZDI} E.
\label{eq:eph}
\end{equation}
where $P_{ZDI}$ is the rotational period used as an input for ZDI and
given in Table~\ref{tab:sample}.

\begin{table*}
 \begin{center}
\caption[]{Journal of observations for AD~Leo. Columns 1--7 list the UT
date, the heliocentric Julian date, the UT time, the observation site,
the exposure time, the peak signal to noise ratio (per 2.6~\kms\
velocity bin) and the rms noise level (relative to the unpolarised
continuum level and per 1.8~\kms\ velocity bin) in the average circular
polarisation profile produced by Least-Squares Deconvolution 
(see Sec.~\ref{sec:obs}). In column 8 we indicate the longitudinal
field computed from Eq.~\ref{eq:bl}. The rotational cycle $E$ from the
ephemeris of Eq.~\ref{eq:eph} is given in column 9. Column 10 lists the
radial velocities (absolute accuracy $0.10~\kms$, internal accuracy
$0.03~\kms$) associated to each exposure.}
  \begin{tabular}{cccccccccc}
\hline
Date & HJD          & UT & Obs. site     & $t_{\rm exp}$ & \sn\ 
& $\sigma_{\rm LSD}$ & $B_{\ell}$ & Cycle & $v_r$ \\
           & (2,453,000+) & (h:m:s) & &   (s)         &       &   (\ptt)
& (G) & & (\kms ) \\
\hline
2007 & & & & & & & \\
Jan 27 & 4127.59748 & 02:14:28 & TBL & 4 $\times$ 600.0 & 274 & 2.6 &
-294.9 $\pm$ 12.9 & 79.999 & 12.40\\
Jan 28 & 4128.60883 & 02:30:45 & TBL & 4 $\times$ 600.0 & 401 & 1.7 &
-233.4 $\pm$ 8.9 & 80.454 & 12.40\\
Jan 29 & 4129.57169 & 01:37:14 & TBL & 4 $\times$ 600.0 & 393 & 1.7 &
-298.2 $\pm$ 10.0 & 80.888 & 12.46\\
Jan 30 & 4130.60841 & 02:30:03 & TBL & 4 $\times$ 600.0 & 472 & 1.4 &
-252.6 $\pm$ 8.0 & 81.355 & 12.36\\
Feb 01 & 4132.59818 & 02:15:14 & TBL & 4 $\times$ 600.0 & 338 & 2.1 &
-252.1 $\pm$ 10.6 & 82.251 & 12.34\\
Feb 02 & 4133.63116 & 03:02:41 & TBL & 4 $\times$ 600.0 & 428 & 1.6 &
-262.5 $\pm$ 8.6 & 82.717 & 12.44\\
Feb 03 & 4134.61119 & 02:33:53 & TBL & 4 $\times$ 600.0 & 395 & 1.7 &
-262.9 $\pm$ 9.3 & 83.158 & 12.34\\
Feb 04 & 4135.62167 & 02:48:56 & TBL & 4 $\times$ 600.0 & 411 & 1.7 &
-238.5 $\pm$ 8.7 & 83.613 & 12.42\\
Feb 05 & 4136.59250 & 02:06:53 & TBL & 4 $\times$ 600.0 & 348 & 2.0 &
-295.9 $\pm$ 10.8 & 84.051 & 12.38\\
\hline
2008 & & & & & & & \\
Jan 19 & 4485.51772 & 00:20:02 & TBL & 4 $\times$ 800.0 & 329 & 2.3 &
-275.1 $\pm$ 11.1 & 241.224 & 12.40\\
Jan 24 & 4489.56829 & 01:32:36 & TBL & 4 $\times$ 600.0 & 398 & 1.8 &
-245.6 $\pm$ 8.8 & 243.049 & 12.40\\
Jan 27 & 4492.53788 & 00:48:39 & TBL & 4 $\times$ 600.0 & 408 & 1.7 &
-284.2 $\pm$ 8.8 & 244.386 & 12.34\\
Jan 28 & 4493.54864 & 01:04:06 & TBL & 4 $\times$ 600.0 & 398 & 1.8 &
-219.1 $\pm$ 8.9 & 244.842 & 12.29\\
Jan 30 & 4495.56109 & 01:21:56 & TBL & 4 $\times$ 600.0 & 341 & 2.2 &
-208.5 $\pm$ 10.1 & 245.748 & 12.32\\
Feb 03 & 4499.56749 & 01:30:58 & TBL & 4 $\times$ 600.0 & 376 & 1.9 &
-259.2 $\pm$ 9.6 & 247.553 & 12.28\\
Feb 05 & 4501.54728 & 01:01:47 & TBL & 4 $\times$ 600.0 & 355 & 2.0 &
-288.3 $\pm$ 10.1 & 248.445 & 12.34\\
Feb 06 & 4502.54747 & 01:02:02 & TBL & 4 $\times$ 600.0 & 414 & 1.7 &
-204.4 $\pm$ 8.1 & 248.895 & 12.33\\
Feb 10 & 4506.55755 & 01:16:25 & TBL & 4 $\times$ 600.0 & 413 & 1.7 &
-224.7 $\pm$ 8.2 & 250.702 & 12.36\\
Feb 12 & 4508.55161 & 01:07:49 & TBL & 4 $\times$ 600.0 & 398 & 1.8 &
-257.9 $\pm$ 8.9 & 251.600 & 12.34\\
Feb 13 & 4509.55640 & 01:14:42 & TBL & 4 $\times$ 600.0 & 398 & 2.2 &
-234.9 $\pm$ 11.7 & 252.052 & 12.40\\
Feb 14 & 4510.55228 & 01:08:45 & TBL & 4 $\times$ 600.0 & 279 & 2.7 &
-281.3 $\pm$ 12.4 & 252.501 & 12.27\\
Feb 15 & 4511.56943 & 01:33:25 & TBL & 4 $\times$ 600.0 & 388 & 1.9 &
-196.0 $\pm$ 8.6 & 252.959 & 12.39\\
Feb 16 & 4512.55367 & 01:10:42 & TBL & 4 $\times$ 600.0 & 405 & 1.7 &
-283.6 $\pm$ 8.8 & 253.403 & 12.36\\
\hline
  \label{tab:log_adleo}
  \end{tabular}
 \end{center}
\end{table*}

\begin{table*}
 \begin{center}
\caption[]{Same as Table~\ref{tab:log_adleo} for EV~Lac.}
  \begin{tabular}{cccccccccc}
\hline
Date & HJD          & UT & Ob. site     & $t_{\rm exp}$ & \sn\ 
& $\sigma_{\rm LSD}$ & $B_{\ell}$ & Cycle & $v_r$ \\
           & (2,453,000+) & (h:m:s) & &   (s)         &       &   (\ptt)
& (G) & & (\kms ) \\
\hline
2006 & & & & & & & &\\
Aug 05 & 3953.07311 & 13:38:30 & CFHT & 4 $\times$ 300.0 & 368 & 2.2 &
-556.8 $\pm$ 17.8 & 0.702 & 0.30\\
Aug 07 & 3955.06634 & 13:28:36 & CFHT & 4 $\times$ 300.0 & 379 & 2.0 &
343.7 $\pm$ 13.3 & 1.157 & 0.14\\
Aug 08 & 3956.06002 & 13:19:26 & CFHT & 4 $\times$ 400.0 & 437 & 1.7 &
-380.1 $\pm$ 12.2 & 1.384 & 0.40\\
Aug 09 & 3957.05969 & 13:18:53 & CFHT & 4 $\times$ 230.0 & 334 & 2.3 &
-464.6 $\pm$ 16.3 & 1.613 & 0.35\\
Aug 10 & 3958.07230 & 13:36:58 & CFHT & 4 $\times$ 250.0 & 332 & 2.4 &
-355.6 $\pm$ 15.2 & 1.844 & 0.38\\
Aug 11 & 3959.07281 & 13:37:39 & CFHT & 4 $\times$ 250.0 & 353 & 2.2 &
297.2 $\pm$ 14.0 & 2.072 & 0.05\\
Aug 12 & 3960.07608 & 13:42:17 & CFHT & 4 $\times$ 250.0 & 329 & 2.5 &
-158.6 $\pm$ 14.4 & 2.302 & 0.33\\
\hline
2007 & & & & & & &\\
% Jul 26 & 4307.56761 & 01:31:32 & TBL & 4 $\times$ 900.0 & 53 & 11.7 &
% -36.1 $\pm$ 125.7 & 81.674 \\
Jul 28 & 4309.54645 & 01:00:53 & TBL & 4 $\times$ 900.0 & 439 & 1.8 &
59.8 $\pm$ 10.5 & 82.126 & 0.30\\
Jul 29 & 4310.56610 & 01:29:05 & TBL & 4 $\times$ 900.0 & 399 & 1.8 &
-421.5 $\pm$ 14.7 & 82.359 & 0.49\\
Jul 30 & 4311.59374 & 02:08:47 & TBL & 4 $\times$ 900.0 & 360 & 2.0 &
-527.4 $\pm$ 17.3 & 82.593 & 0.60\\
Jul 31 & 4312.59372 & 02:08:40 & TBL & 4 $\times$ 600.0 & 326 & 2.5 &
34.2 $\pm$ 14.4 & 82.822 & 0.22\\
Aug 01 & 4313.59576 & 02:11:31 & TBL & 4 $\times$ 600.0 & 281 & 3.0 &
267.6 $\pm$ 18.5 & 83.051 & 0.30\\
Aug 03 & 4315.60183 & 02:20:05 & TBL & 4 $\times$ 600.0 & 306 & 2.5 &
-481.7 $\pm$ 18.8 & 83.509 & 0.62\\
Aug 04 & 4316.59985 & 02:17:09 & TBL & 4 $\times$ 600.0 & 330 & 2.4 &
-271.4 $\pm$ 15.5 & 83.737 & 0.29\\
Aug 05 & 4317.67118 & 03:59:47 & TBL & 4 $\times$ 600.0 & 273 & 3.0 &
338.1 $\pm$ 19.1 & 83.982 & 0.46\\
% Aug 09 & 4321.57247 & 01:37:20 & TBL & 4 $\times$ 600.0 & 50 & 13.0 &
% -194.7 $\pm$ 245.9 & 84.873 \\
Aug 10 & 4322.59520 & 02:09:60 & TBL & 4 $\times$ 600.0 & 303 & 2.7 &
107.0 $\pm$ 16.1 & 85.106 & 0.26\\
Aug 11 & 4323.59772 & 02:13:34 & TBL & 4 $\times$ 600.0 & 235 & 3.5 &
-353.7 $\pm$ 21.1 & 85.335 & 0.47\\
Aug 15 & 4327.58824 & 01:59:40 & TBL & 4 $\times$ 600.0 & 301 & 2.5 &
-318.8 $\pm$ 16.8 & 86.247 & 0.30\\
Aug 18 & 4330.58129 & 01:49:29 & TBL & 4 $\times$ 600.0 & 308 & 2.4 &
378.2 $\pm$ 17.0 & 86.930 & 0.29\\
Aug 19 & 4331.51487 & 00:13:47 & TBL & 4 $\times$ 600.0 & 339 & 2.3 &
-62.2 $\pm$ 13.8 & 87.144 & 0.21\\
% Aug 26 & 4339.48187 & 23:25:56 & TBL & 4 $\times$ 600.0 & 66 & 16.0 &
% 388.5 $\pm$ 91.5 & 88.963 \\
Aug 28 & 4340.53002 & 00:35:14 & TBL & 4 $\times$ 600.0 & 279 & 2.8 &
-235.8 $\pm$ 17.8 & 89.203 & 0.30\\
Aug 31 & 4343.52117 & 00:22:25 & TBL & 4 $\times$ 600.0 & 258 & 3.1 &
232.7 $\pm$ 18.6 & 89.886 & 0.30\\
\hline
  \label{tab:log_evlac}
  \end{tabular}
 \end{center}
\end{table*}

\begin{table*}
 \begin{center}
\caption[]{Same as Table~\ref{tab:log_adleo} for YZ~CMi.}
  \begin{tabular}{cccccccccc}
\hline
Date & HJD          & UT & Ob. site     & $t_{\rm exp}$ & \sn\ 
& $\sigma_{\rm LSD}$ & $B_{\ell}$ & Cycle & $v_r$  \\
           & (2,453,000+) & (h:m:s) & &   (s)         &       &   (\ptt)
& (G) & & (\kms ) \\
\hline
2007 & & & & & & &\\
Jan 26 & 4127.43869 & 22:24:17 & TBL & 4 $\times$ 900.0 & 235 & 3.9 &
-401.9 $\pm$ 31.4 & 64.057 & 26.66\\
Jan 27 & 4128.47944 & 23:22:59 & TBL & 4 $\times$ 900.0 & 255 & 3.3 &
-782.5 $\pm$ 33.3 & 64.433 & 26.60\\
Jan 29 & 4130.47395 & 23:15:08 & TBL & 4 $\times$ 900.0 & 324 & 2.6 &
-520.4 $\pm$ 25.3 & 65.153 & 26.74\\
Feb 01 & 4133.50014 & 23:52:56 & TBL & 4 $\times$ 900.0 & 254 & 3.7 &
-710.2 $\pm$ 40.3 & 66.246 & 26.82\\
Feb 03 & 4135.49442 & 23:44:47 & TBL & 4 $\times$ 900.0 & 280 & 3.3 &
-156.7 $\pm$ 25.1 & 66.965 & 26.51\\
Feb 04 & 4136.46196 & 22:58:05 & TBL & 4 $\times$ 900.0 & 261 & 3.4 &
-781.5 $\pm$ 36.3 & 67.315 & 26.66\\
Feb 08 & 4140.47749 & 23:20:37 & TBL & 4 $\times$ 900.0 & 260 & 3.8 &
-62.3 $\pm$ 27.9 & 68.764 & 26.44\\
\hline
Dec 28 & 4462.62633 & 02:54:53 & TBL & 4 $\times$ 1200.0 & 289 & 3.9 &
-279.8 $\pm$ 28.6 & 185.064 & 26.28\\
Dec 29 & 4463.65629 & 03:37:59 & TBL & 4 $\times$ 1200.0 & 323 & 3.0 &
-560.9 $\pm$ 27.0 & 185.435 & 26.75\\
Dec 31 & 4465.67053 & 03:58:24 & TBL & 4 $\times$ 900.0 & 238 & 4.4 &
-166.5 $\pm$ 30.2 & 186.163 & 26.43\\
2008 & & & & & & &\\
Jan 01 & 4466.66384 & 03:48:43 & TBL & 4 $\times$ 1100.0 & 305 & 3.0 &
-680.3 $\pm$ 29.4 & 186.521 & 26.59\\
% Jan 02 & 4467.64822 & 03:26:12 & TBL &  2200.0 s (1100.0+1100.0 s)
%  & 144 & 7.2 & -430.2 $\pm$ 50.2 & 186.877 \\
Jan 03 & 4468.66111 & 03:44:43 & TBL & 4 $\times$ 1100.0 & 272 & 3.7 &
-97.7 $\pm$ 26.3 & 187.242 & 26.69\\
Jan 20 & 4485.93731 & 10:22:12 & CFHT & 4 $\times$ 220.0 & 240 & 3.9 &
-599.6 $\pm$ 31.0 & 193.479 & 26.65\\
% Jan 21 & 4487.46385 & 23:00:26 & TBL & 4 $\times$ 1200.0 & 311 & 4.5 &
% -1089.5 $\pm$ 150.8 & 194.030 \\
% Jan 22 & 4488.52001 & 24:21:18 & TBL & 4 $\times$ 1200.0 & 281 & 3.3 &
% -452.3 $\pm$ 26.7 & 194.412 \\
Jan 23 & 4488.52001 & 00:21:18 & TBL & 4 $\times$ 1200.0 & 281 & 3.3 &
-452.3 $\pm$ 26.7 & 194.412 & 26.76\\
Jan 23 & 4489.45108 & 22:42:04 & TBL & 4 $\times$ 1200.0 & 252 & 3.8 &
-624.4 $\pm$ 32.0 & 194.748 & 26.84\\
Jan 24 & 4490.53391 & 00:41:21 & TBL & 4 $\times$ 1200.0 & 290 & 3.4 &
-258.2 $\pm$ 24.8 & 195.139 & 26.34\\
Jan 25 & 4491.46536 & 23:02:40 & TBL & 4 $\times$ 1200.0 & 254 & 3.9 &
-575.7 $\pm$ 31.2 & 195.475 & 26.68\\
Jan 26 & 4492.45361 & 22:45:46 & TBL & 4 $\times$ 1200.0 & 317 & 3.1 &
-538.7 $\pm$ 27.4 & 195.832 & 26.82\\
Jan 27 & 4493.46567 & 23:03:09 & TBL & 4 $\times$ 1200.0 & 324 & 3.2 &
-116.5 $\pm$ 22.8 & 196.197 & 26.46\\
Jan 28 & 4494.53067 & 00:36:46 & TBL & 4 $\times$ 1200.0 & 260 & 3.8 &
-724.1 $\pm$ 35.0 & 196.581 & 26.57\\
Jan 29 & 4495.47886 & 23:22:12 & TBL & 4 $\times$ 1200.0 & 294 & 3.6 &
-537.6 $\pm$ 32.6 & 196.924 & 26.37\\
Feb 02 & 4499.47856 & 23:21:54 & TBL & 4 $\times$ 1200.0 & 281 & 3.7 &
-292.8 $\pm$ 26.5 & 198.368 & 26.74\\
Feb 04 & 4501.45937 & 22:54:20 & TBL & 4 $\times$ 1200.0 & 217 & 5.0 &
-410.1 $\pm$ 36.2 & 199.083 & 26.15\\
Feb 05 & 4502.46143 & 22:57:21 & TBL & 4 $\times$ 1200.0 & 299 & 3.4 &
-456.3 $\pm$ 28.4 & 199.445 & 26.66\\
Feb 06 & 4503.49639 & 23:47:44 & TBL & 4 $\times$ 1200.0 & 316 & 3.1 &
-533.7 $\pm$ 26.6 & 199.818 & 26.80\\
Feb 09 & 4506.46767 & 23:06:31 & TBL & 4 $\times$ 1200.0 & 124 & 8.4 &
-491.3 $\pm$ 58.9 & 200.891 & 26.71\\
Feb 11 & 4508.46975 & 23:09:38 & TBL & 4 $\times$ 1200.0 & 282 & 3.2 &
-686.2 $\pm$ 29.7 & 201.614 & 26.60\\
Feb 12 & 4509.47423 & 23:16:08 & TBL & 4 $\times$ 1200.0 & 317 & 3.2 &
-485.5 $\pm$ 26.4 & 201.976 & 26.49\\
Feb 13 & 4510.46991 & 23:09:58 & TBL & 4 $\times$ 1200.0 & 268 & 3.7 &
-165.3 $\pm$ 25.5 & 202.336 & 26.76\\
Feb 14 & 4511.48644 & 23:33:50 & TBL & 4 $\times$ 1200.0 & 280 & 3.5 &
-653.1 $\pm$ 30.6 & 202.703 & 26.74\\
Feb 15 & 4512.47158 & 23:12:30 & TBL & 4 $\times$ 1200.0 & 320 & 3.2 &
-421.7 $\pm$ 25.6 & 203.058 & 26.17\\
Feb 16 & 4513.47114 & 23:11:56 & TBL & 4 $\times$ 1200.0 & 326 & 3.0 &
-351.3 $\pm$ 23.6 & 203.419 & 26.79\\

\hline
  \label{tab:log_yzcmi}
  \end{tabular}
 \end{center}
\end{table*}

\begin{table*}
 \begin{center}
\caption[]{Same as Table~\ref{tab:log_adleo} for EQ~Peg~A.}
  \begin{tabular}{cccccccccc}
\hline
Date & HJD          & UT & Ob. site     & $t_{\rm exp}$ & \sn\ 
& $\sigma_{\rm LSD}$ & $B_{\ell}$ & Cycle & $v_r$ \\
           & (2,453,000+) & (h:m:s) & &   (s)         &       &   (\ptt)
& (G) & & (\kms ) \\
\hline
2006 & & & & & & & &\\
Aug 05 & 3952.99538 & 11:46:33 & CFHT & 4 $\times$ 200.0 & 322 & 2.5 &
264.9 $\pm$ 18.6 & 2.799 & 0.39\\
Aug 05 & 3953.11181 & 14:34:13 & CFHT & 4 $\times$ 160.0 & 295 & 2.9 &
191.0 $\pm$ 20.4 & 2.908 & 0.78\\
Aug 07 & 3954.97268 & 11:13:40 & CFHT & 4 $\times$ 200.0 & 327 & 2.5 &
409.9 $\pm$ 19.4 & 4.647 & 0.27\\
Aug 07 & 3955.13847 & 15:12:24 & CFHT & 4 $\times$ 200.0 & 323 & 2.6 &
279.0 $\pm$ 18.8 & 4.802 & 0.47\\
Aug 08 & 3955.98480 & 11:31:02 & CFHT & 4 $\times$ 160.0 & 283 & 3.0 &
412.4 $\pm$ 22.3 & 5.593 & 0.31\\
Aug 08 & 3956.14334 & 15:19:20 & CFHT & 4 $\times$ 300.0 & 374 & 2.1 &
326.9 $\pm$ 16.2 & 5.741 & 0.34\\
Aug 09 & 3956.99077 & 11:39:32 & CFHT & 4 $\times$ 160.0 & 261 & 3.2 &
427.8 $\pm$ 23.6 & 6.533 & 0.28\\
Aug 09 & 3957.12529 & 14:53:14 & CFHT & 4 $\times$ 160.0 & 289 & 2.9 &
412.4 $\pm$ 21.7 & 6.659 & 0.31\\
Aug 09 & 3957.13682 & 15:09:50 & CFHT & 4 $\times$ 160.0 & 284 & 2.9 &
409.9 $\pm$ 21.8 & 6.670 & 0.30\\
Aug 10 & 3957.98897 & 11:36:51 & CFHT & 4 $\times$ 160.0 & 296 & 2.9 &
427.8 $\pm$ 21.9 & 7.466 & 0.21\\
Aug 10 & 3958.14147 & 15:16:26 & CFHT & 4 $\times$ 160.0 & 273 & 3.1 &
460.5 $\pm$ 23.1 & 7.609 & 0.34\\
Aug 11 & 3958.99132 & 11:40:09 & CFHT & 4 $\times$ 160.0 & 290 & 3.1 &
379.4 $\pm$ 22.3 & 8.403 & 0.10\\
Aug 11 & 3959.13871 & 15:12:23 & CFHT & 4 $\times$ 160.0 & 274 & 3.1 &
433.1 $\pm$ 23.1 & 8.541 & 0.32\\
Aug 12 & 3959.99595 & 11:46:44 & CFHT & 4 $\times$ 160.0 & 272 & 3.4 &
379.3 $\pm$ 24.3 & 9.342 & -0.04\\
Aug 12 & 3960.14401 & 15:19:55 & CFHT & 4 $\times$ 160.0 & 262 & 3.5 &
382.2 $\pm$ 24.8 & 9.480 & 0.25\\
\hline
  \label{tab:log_eqpega}
  \end{tabular}
 \end{center}
\end{table*}

\begin{table*}
 \begin{center}
\caption[]{Same as Table~\ref{tab:log_adleo} for EQ~Peg~B.}
  \begin{tabular}{cccccccccc}
\hline
Date & HJD          & UT & Ob. site     & $t_{\rm exp}$ & \sn\ 
& $\sigma_{\rm LSD}$ & $B_{\ell}$ & Cycle & $v_r$ \\
           & (2,453,000+) & (h:m:s) & &   (s)         &       &   (\ptt)
& (G) & & (\kms ) \\
\hline
Aug 05 & 3953.01352 & 12:12:41 & CFHT & 4 $\times$ 320.0 & 194 & 5.5 &
315.4 $\pm$ 40.8 & 4.532 & 3.32\\
Aug 05 & 3953.09585 & 14:11:14 & CFHT & 4 $\times$ 320.0 & 194 & 5.4 &
403.9 $\pm$ 41.4 & 4.655 & 3.32\\
Aug 07 & 3955.00502 & 12:00:15 & CFHT & 4 $\times$ 320.0 & 194 & 5.5 &
296.5 $\pm$ 40.5 & 7.526 & 3.31\\
Aug 07 & 3955.12027 & 14:46:12 & CFHT & 4 $\times$ 320.0 & 187 & 5.6 &
358.5 $\pm$ 41.5 & 7.700 & 3.41\\
Aug 08 & 3956.00013 & 11:53:06 & CFHT & 4 $\times$ 300.0 & 187 & 5.8 &
243.4 $\pm$ 42.3 & 9.023 & 3.48\\
Aug 08 & 3956.12205 & 14:48:40 & CFHT & 4 $\times$ 400.0 & 222 & 4.7 &
197.8 $\pm$ 35.7 & 9.206 & 3.10\\
Aug 09 & 3957.00746 & 12:03:34 & CFHT & 4 $\times$ 280.0 & 178 & 6.0 &
278.2 $\pm$ 44.1 & 10.538 & 3.45\\
Aug 09 & 3957.10933 & 14:30:16 & CFHT & 4 $\times$ 280.0 & 180 & 6.0 &
376.4 $\pm$ 44.5 & 10.691 & 3.26\\
Aug 10 & 3958.00771 & 12:03:51 & CFHT & 4 $\times$ 300.0 & 176 & 6.1 &
204.8 $\pm$ 44.0 & 12.042 & 3.51\\
Aug 10 & 3958.12127 & 14:47:22 & CFHT & 4 $\times$ 300.0 & 178 & 6.0 &
205.1 $\pm$ 43.5 & 12.212 & 3.16\\
Aug 11 & 3959.00904 & 12:05:40 & CFHT & 4 $\times$ 300.0 & 150 & 7.4 &
287.3 $\pm$ 52.7 & 13.547 & 3.38\\
% Aug 11 & 3959.12639 & 14:54:39 & CFHT &  423.1 s (300.0+102.9+18.2+2.0
% s) & 33 & 23.0 & 394.5 $\pm$ 148.4 & 13.724 \\
% Aug 11 & 3959.15537 & 15:36:22 & CFHT &  903.7 s (300.0+300.0+300.0+3.7
% s) & 42 & 18.2 & 56.2 $\pm$ 169.7 & 13.767 \\
Aug 12 & 3960.01246 & 12:10:30 & CFHT & 4 $\times$ 300.0 & 152 & 7.2 &
222.1 $\pm$ 50.4 & 15.056 & 3.62\\
Aug 12 & 3960.12839 & 14:57:26 & CFHT & 4 $\times$ 300.0 & 155 & 7.4 &
299.5 $\pm$ 51.0 & 15.231 & 3.08\\
\hline
  \label{tab:log_eqpegb}
  \end{tabular}
 \end{center}
\end{table*}

\section{Model description}
\label{sec:mod}
For each star of our sample, our aim is to infer the topology of the
surface magnetic field from the time series of circularly polarised
(Stokes $V$) LSD profiles we obtained. This can be achieved using a
tomographic imaging code. In this part we briefly present the main
features of our imaging code, the physical model used to describe
the Stokes $I$ and $V$ line profiles, and the way we use this code to
provide constraints on rotational period and differential rotation.
 \subsection{Zeeman-Doppler Imaging (ZDI)}
 \label{sec:mod-zdi}
Circularly polarised light emitted by a star informs us about the
longitudinal magnetic field at its surface. Thanks to the Doppler
effect, magnetic regions at the surface of a rapidly rotating star
produce Stokes $V$ signatures whose wavelength strongly correlates with
their spatial position; in this respect a circularly polarised line
profile can be seen as 1D image of the longitudinal magnetic field. By
analysing how these signatures are modulated by rotation, it is possible
to reconstruct a 2D map of the surface magnetic field. See
\cite{Brown91} and \cite{Donati97b} for more details about ZDI and its
performances. As we demonstrate in this paper, and was already  shown by
\cite{Donati06b} for $\tau$ Sco ($\vsini \simeq 5~\kms$), even for
slowly rotating stars ZDI is able to recover some information about the
large-scale surface magnetic field. In all cases, we need to set
$\ell>6$ to be able to reproduce rotational modulation in our data.

The ZDI code we employ here is based on a spherical harmonics
description of each component of the magnetic field vector, implemented
by \cite{Donati06b}. Compared with the conventional ZDI technique (which
described the field as a set of independent values), this approach
allows us to reconstruct a physically meaningful magnetic field as the
sum of a poloidal field and a toroidal field \citep{Chandra61}. Such a
decomposition is of obvious interest for all studies on stellar dynamos.
Moreover, this method proved to be more efficient than the old one at
recovering simple low order topologies such as dipoles, even from Stokes
$V$ data sets only \citep{Donati01}.

ZDI works by comparing observational data to synthetic spectra computed
from a guess magnetic map. The map is iteratively updated until the
corresponding spectra fit the observations within a given \chisq\
level. In order to compute the synthetic spectra, the surface of the
star is divided into a grid of $\sim1,000$ cells on which the magnetic
field components are computed from the coefficients of the spherical
harmonics expansion. The contribution of each individual pixel is
computed from a model based on Unno-Rachkovsky's equations (see
Sec.~\ref{sec:mod-line}).

Given the projected rotational velocities for our sample ($\vsini <
30~\kms$) and considering the local profile width ($\simeq 9~\kms$,
M08), we infer that the maximum number of spatially resolved elements
across the equator is about 20. Therefore, using a grid of 1,000 cells
at the surface of the star (the equatorial ring of the grid is made of
about 70 elements, depending on the inclination of the star) is
perfectly adequate for our needs.

As the inversion problem is partly ill-posed, several magnetic
topologies can fit a set of observations, for a given \chisq\ level.
Optimal reconstruction is achieved by choosing the maximum entropy
solution, i.e. the one which contains the least informational content
\citep{Skilling84}. We chose here a quadratic form for the entropy : 

\begin{equation}
 S = -\sum_{\ell, m} \ell \left(\alpha_{\ell, m}^2 + \beta_{\ell, m}^2 +
\gamma_{\ell, m}^2\right)
\label{eq:ent}
\end{equation}
where $\alpha_{\ell, m}$, $\beta_{\ell, m}$, $\gamma_{\ell, m}$ are the
spherical harmonics coefficient of order $(\ell, m)$ describing
respectively the radial, orthoradial poloidal and toroidal field
components \citep[see ][ for more details]{Donati06b}. This
functional, one of the simplest possible forms, is well-suited for
magnetic fields reconstruction since it allows for negative values (as
opposed to the conventional expression of the Shannon entropy). 

 \subsection{Modelling of the local line profiles}
 \label{sec:mod-line}
As explained in Sec.~\ref{sec:mod-zdi}, the local Stokes $I$ and $V$
line profiles are computed from a simple model based on
Unno-Rachkovsky's equations \citep[][]{Unno56}, similar to that used by
\cite{Donati08}.
We add two degrees of freedom to the Unno-Rachkovsky's model, the
filling factors $f_I$ and $f_V$:
\begin{equation}
\left\lbrace
\begin{array}{l}
I = f_I \times \IUR(\lambda_B) + (1-f_I) \times \Iq \\
V = f_V \times \VUR(\lambda_B) \\
\lambda_B = 4.67 \times 10^{-12} \,g_{\rm eff}\,
\lambda_0^2 \, B/f_V \\
\end{array}
\right.
\end{equation}
where \IUR\ and \VUR\ are the Stokes parameters from Unno-Rachkovsky's
equations (see \citealt{Landi92} for more details), \Iq\ is Stokes $I$
computed without magnetic field, $\lambda_B$ is the Zeeman splitting (in
nm), $\lambda_0$ and $g_{\rm eff}$  are respectively central wavelength
(in nm) and the averaged effective Land\'e factor of the synthetic LSD
line, and $B$ is the longitudinal magnetic flux expressed in Gauss.

With this model, we assume that each grid cell is uniformly covered by a
fraction $f_I$ of magnetic regions \citep[e.g.,][]{Saar88} and a
fraction $f_V$ of magnetic regions producing a net circularly polarised
signature (and thus a fraction $f_I-f_V$ of magnetic regions producing,
on the average, no circularly polarised signature). We justify the use
of two different filling factors by the fact that Stokes $I$ and $V$ are
not affected in the same way by magnetic fields. In particular
signatures corresponding to small bipolar regions of magnetic field
cancel each other in circular polarisation whereas they add up in
unpolarised spectra. We further assume that both $f_I$ and $f_V$ have a
constant value over the stellar surface. 

The filling factor $f_V$ is well constrained by our observations, except
for the fastest rotators. It allows us to reconcile the discrepancy
between the amplitude of Stokes $V$ signatures (constrained by the
magnetic flux $B$) and the Zeeman splitting observed in Stokes $V$
profiles (constrained by the magnetic field strength $B/f_V$). Since
$f_I$ is partly degenerate with other line parameters, we only find a
coarse estimate. Values of $f_I$ around 0.5 allow us to match the observed
Stokes I profiles. Setting $f_I=1.0$ results in a large variability in
synthetic Stokes $I$ profiles that is not observed. Recovered $f_I$ are
typically 3 to 5 times larger than $f_V$, this is roughly consistent with
the ratio of the magnetic fluxes reported here and by \cite{Reiners07}.

We further assume that continuum limb-darkening varies linearly with the
cosine of the limb angle (with a slope of $u=0.6964$,
\citealt{Claret04}). Using a quadratic (rather than linear)
dependence produces no visible change in the result.

 \subsection{Modelling of differential rotation}
 \label{sec:mod-diffrot}

In order to reconstruct a magnetic topology from a time-series of
Stokes $V$ spectra, the ZDI code requires the rotation period of the
observed star as an input. The inversion procedure being quite sensitive
to the assumed period, ZDI can provide a strong constraint on this
parameter. The period resulting in the minimum \chisqr\ at a
given informational content (i.e. a given averaged magnetic flux value)
is the most probable. This is how \Prot\ are derived in this paper.

Differential rotation can be measured as well by proceeding
as in \cite{Petit02} and M08. We assume that the latitudinal variation
of rotation rate can be expressed as:
\begin{equation}
\Omega(\theta) = \Omeq - \dOm \sin^2 \theta
\label{eq:drot}
\end{equation}
where $\Omega_{\rm eq}$ is the rotation rate at the equator and $d\Omega$
the difference in rotation rate between the equator and the pole. This
law is used to compute the phase shift of each ring of the  grid at
any observation epoch with respect to its position at a reference epoch.
Each synthetic Stokes $V$ spectrum (see Sec. \ref{sec:mod-line}) is then
computed from the magnetic field distribution at the reference epochs
distorted by the aforementioned phase shifts.

For a set of pairs (\Omeq;\dOm) within a reasonable range of values, we
run ZDI and derive the corresponding  magnetic map along with the
associated \chisqr\ level. By fitting a paraboloid to the \chisqr\ surface
derived in this process \citep{Donati03b}, we can easily infer the
magnetic
topology that yields the best fit to the data along with the corresponding
differential rotation parameters and error bars.

\section{AD~Leo = GJ~388}
\label{sec:adleo}

\begin{figure*}
 \center{%
 \includegraphics[scale=0.45]{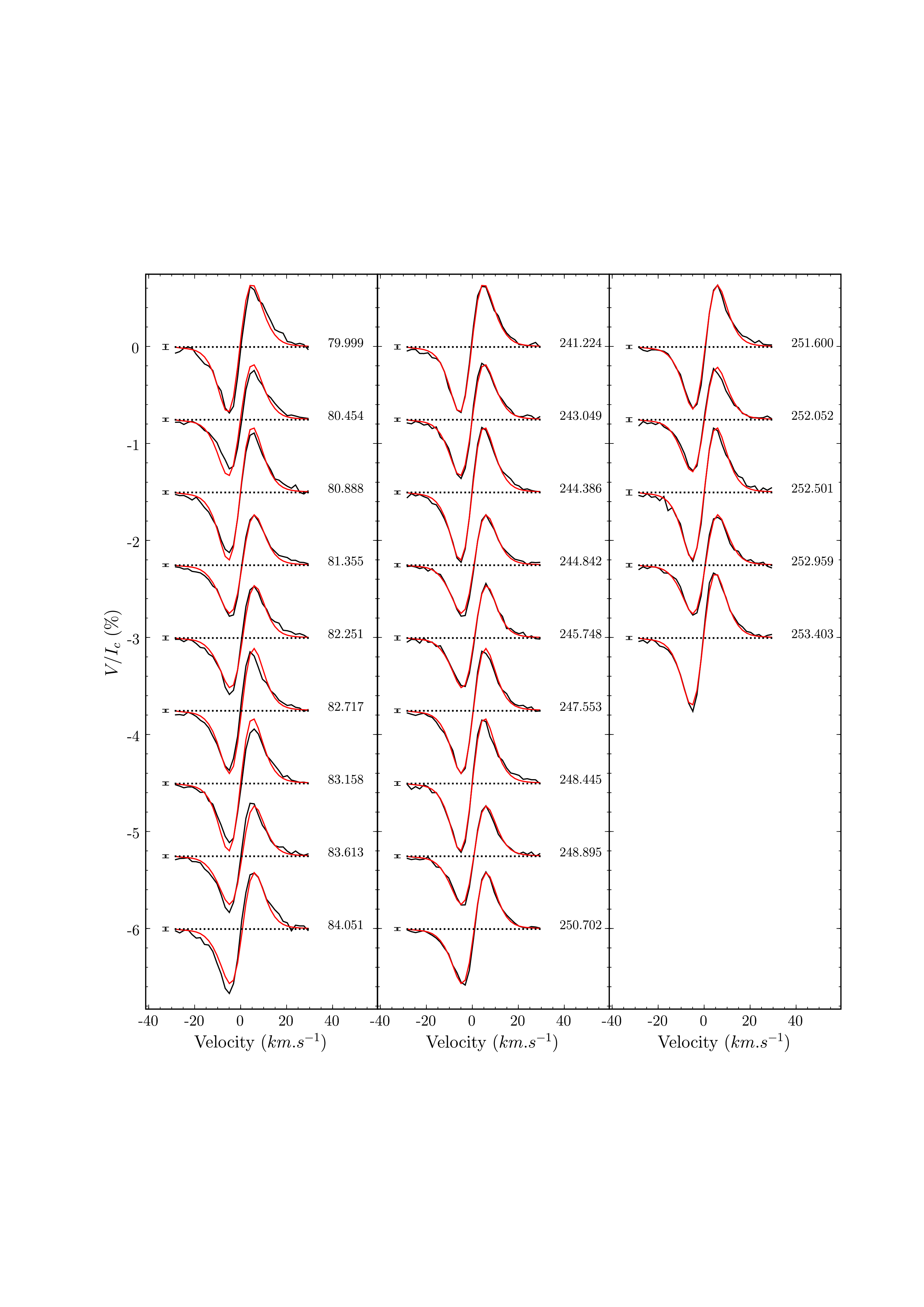}
 \caption[]{Time-series of Stokes $V$ profiles of AD~Leo, in the
rest-frame of the star, from our 2007 (left-hand column) and 2008
(middle and right-hand columns) data sets. Synthetic profiles
corresponding to our magnetic models (red lines) are superimposed to the
observed LSD profiles (black lines). Left to each profile a
$\pm1-\sigma$ error bar is shown. The rotational phase and cycle of each
observation is also mentioned right to each profile. Successive profiles
are shifted vertically for clarity purposes and the associated reference
levels ($V=0$) are plotted as dotted lines.}
 \label{fig:zdi_spec_adleo}}
\end{figure*}

\begin{figure*}
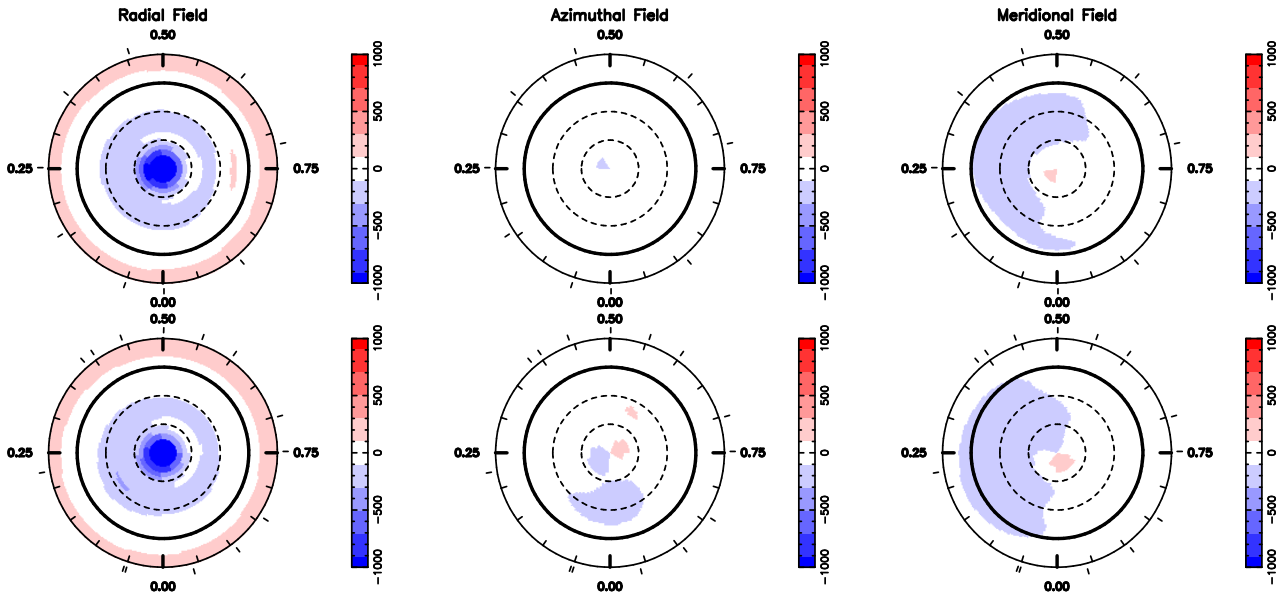

 \center{%
 \includegraphics[scale=0.9]{fig/adleo_V_07g_map.ps}
 \includegraphics[scale=0.9]{fig/adleo_V_08g_map.ps}
 \caption[]{Surface magnetic flux of AD~Leo as derived from our 2007
(upper row) and 2008 (lower row) data sets. The three components of the
field in spherical coordinates are displayed from left to right (flux
values labelled in G).  The star is shown in flattened polar projection
down to latitudes of $-30\degr$, with the equator depicted as a bold
circle and parallels as dashed circles. Radial ticks around each plot
indicate phases of observations.}
 \label{fig:zdi_maps_adleo} 
}
\end{figure*}

\begin{figure}
 \center{%
 \includegraphics[scale=0.4]{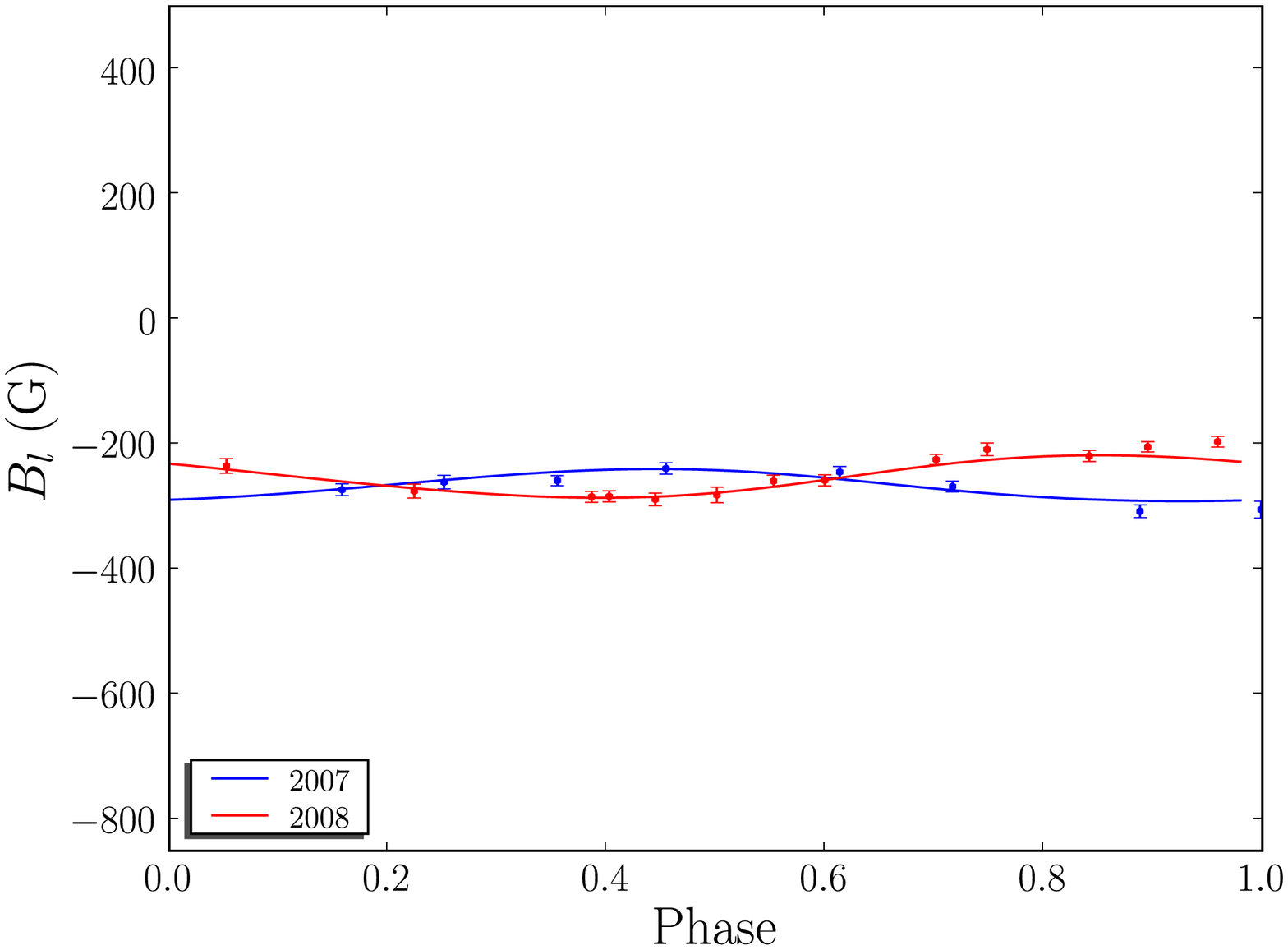}
 \caption[]{Longitudinal magnetic field of AD~Leo as computed from the
observed LSD Stokes $I$ and $V$ profiles for each observation epochs,
1-$\sigma$ error bars are also plotted (see Tab~\ref{tab:log_adleo}).
The solid lines represent the longitudinal field corresponding to the
magnetic topologies reconstructed by ZDI and shown in
Fig~\ref{fig:zdi_maps_adleo}. The scale is the same for all the plots of
longitudinal field.}
 \label{fig:bl_adleo}}
\end{figure}

We observed AD~Leo in January-February 2007, and then one year later
in January-February 2008 (see Tab.~\ref{tab:log_adleo}). We 
respectively secured 9 and 14 spectra at each epoch (see
Fig.~\ref{fig:zdi_spec_adleo}) providing complete though not very dense
coverage of the rotational cycle (see Fig.~\ref{fig:zdi_maps_adleo}).
Both time-series are very similar, we detect a strong signature of
negative polarity (i.e. longitudinal field directed towards the star)
exhibiting only very weak time-modulation (see Fig.~\ref{fig:bl_adleo}).
We thus expect that the star is seen nearly pole-on. We measure mean RV
of $12.39~\kms$ and $12.35~\kms$ in 2007 and 2008 respectively, in good
agreement with the value reported by \cite{Nidever02} of
$12.42\pm0.1~\kms$. The dispersion about these mean RV is equal to
$0.04~\kms$ at both epochs, i.e. close to the internal RV accuracy of
NARVAL (about $0.03~\kms$, see Sec.~\ref{sec:obs}). These variations
likely reflect the internal RV jitter of AD~Leo since we observe a
smooth variation of RV as a function of the rotational phase (even
for observations occurring at different rotation cycles). We notice
that RV and $B_l$ are in quadrature at both epochs. Given the previously
reported stellar parameters $\vsini = 3.0~\kms$ \citep{Reiners07}, a
rotation period of $2.7~\d$ \citep[][]{Spiesman86} and $\rstar \leq
0.40~\rsun$ (see Tab.~\ref{tab:sample}) we indeed infer $i\simeq 20\degr$.

We first process separately the 2007 and 2008 data described above with
ZDI assuming $\vsini = 3.0~\kms$, $i = 20\degr$, and reconstruct modes 
up to order $\ell=8$, which is enough given the low rotational velocity
of AD~Leo. It is possible to fit the Stokes $V$ spectra down to 
$\chisqr = 2.0$ (from an initial $\chisqr \simeq 250$) for both data  
sets if we assume $P_{ZDI} = 2.22~\d$, which is significantly lower than
the formerly estimated photometric period. Very similar results are
obtained whether we assume that the field is purely poloidal or the
presence a toroidal component. In the latter case toroidal fields only
account for 5\% of the overall recovered magnetic energy in 2008,
whereas they are only marginally recovered from the 2007 -- sparser --
dataset (1\%).

Very similar large-scale magnetic fields are recovered from both
data sets (see Fig.\ref{fig:zdi_maps_adleo}), with an average recovered
magnetic flux $B \simeq 0.2~\kG$. We report a strong polar spot
of radial field of maximum magnetic flux $B = 1.3~\kG$ as the
dominant feature of the surface magnetic field. The spherical harmonics
decomposition of the surface magnetic field confirms what can be
inferred from the magnetic maps. First, the prominent mode is the
radial component of a dipole aligned with the rotational axis i.e. the
$\ell=1, m=0$ mode of the radial component ($\alpha(1;0)$ contains
more than 50\% of the reconstructed magnetic energy). Secondly, the
magnetic topology is strongly axisymmetric with about 90\% of the
energy in $m=0$ modes. Thirdly, among the recovered modes the lower
order ones encompass most of the reconstructed magnetic energy ($\simeq
60\%$ in the dipole modes, i.e. modes $\alpha$ or $\beta$ modes of order
$\ell=0$), though we cannot fit our data down to $\chisqr=2.0$ if we do
not include modes up to order $\ell=8$.

We use ZDI to measure differential rotation as explained in
Section~\ref{sec:mod-diffrot}. The \chisqr\ map resulting from the
analysis of the 2008 dataset does not features a clear paraboloid but
rather a long valley with no well-defined minimum. If we assume
solid-body rotation, a clear minimum is obtained at $\Prot = 2.24 \pm
0.02~\d$ (3-$\sigma$ error-bar).

To estimate the degree at which the magnetic topology remained stable
over 1~yr, we merge our 2007 and 2008 data sets together and try to fit
them simultaneously with a single field structure. Assuming rigid-body
rotation, it is possible to fit the complete data set down to
$\chisqr=2.4$, demonstrating that intrinsic variability between
January 2007 and January 2008 is detectable in our data though very
limited. The corresponding rotation period is $\Prot=2.2399\pm0.0006~\d$
(3-$\sigma$ error-bar). We also find aliases for both shorter and longer
periods, corresponding to shifts of $\sim0.014~\d$. The nearest local
minima located at $\Prot=2.2264~\d$ and $\Prot=2.2537~\d$, are
associated with $\Delta\chisq$ values of 36 and 31 respectively; the
corresponding rotation rates are thus fairly excluded. The periods we
find for the 2008 data set alone or for both data sets are compatible
with each other. But they are not with the period reported by
\cite{Spiesman86} ($2.7\pm0.05~\d$) based on 9 photometric measurements,
for which we believe that the error bar was underestimated.

\section{EV~Lac = GJ~873 = HIP~112460}
\label{sec:evlac}

\begin{figure}
 \center{%
 \includegraphics[scale=0.5]{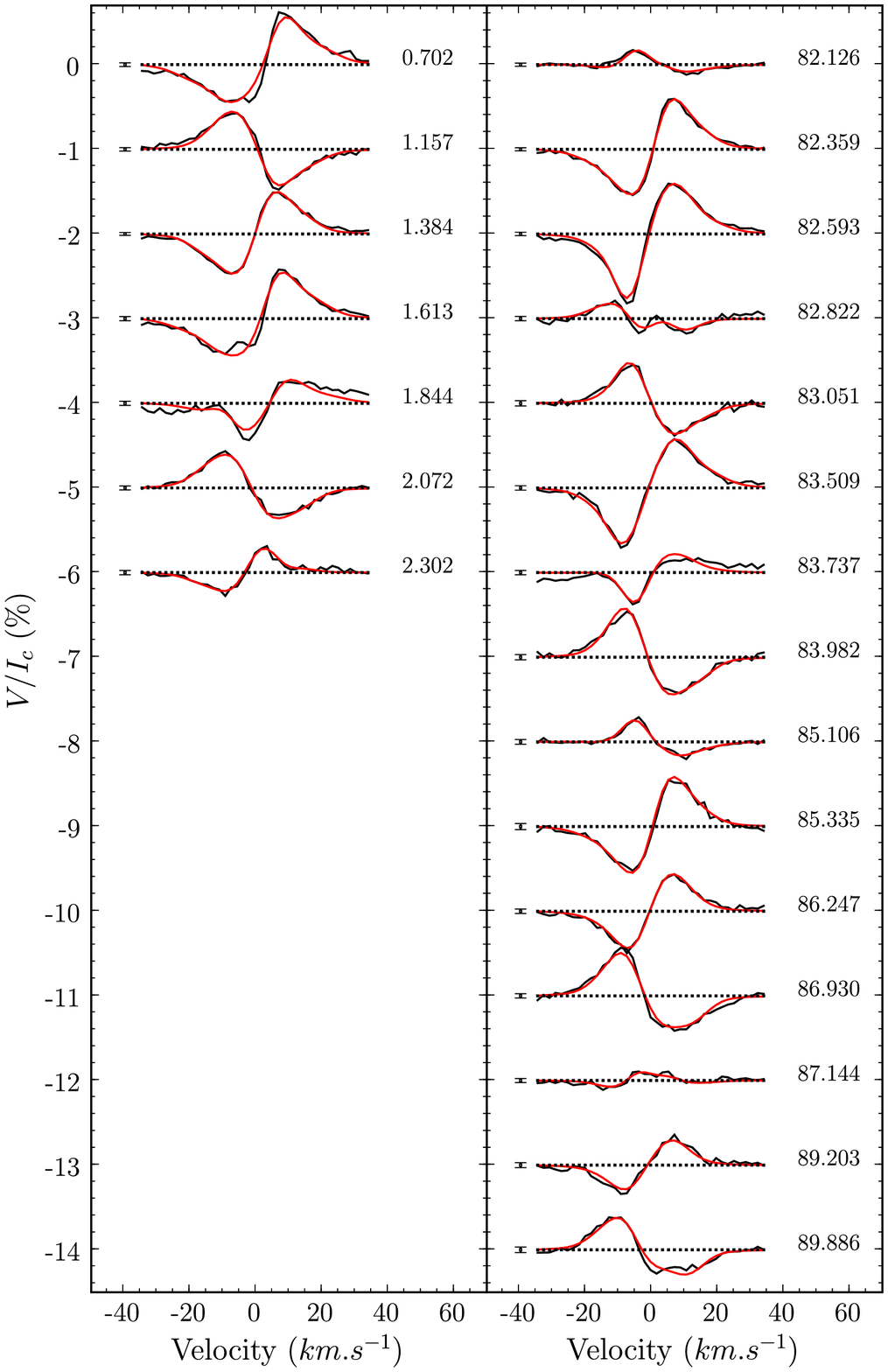}
 \caption[]{Same as Fig.~\ref{fig:zdi_spec_adleo} for EV~Lac 2006
(left-hand column) and 2007 (right-hand column) data sets.}
 \label{fig:zdi_spec_evlac}}
\end{figure}

\begin{figure*}
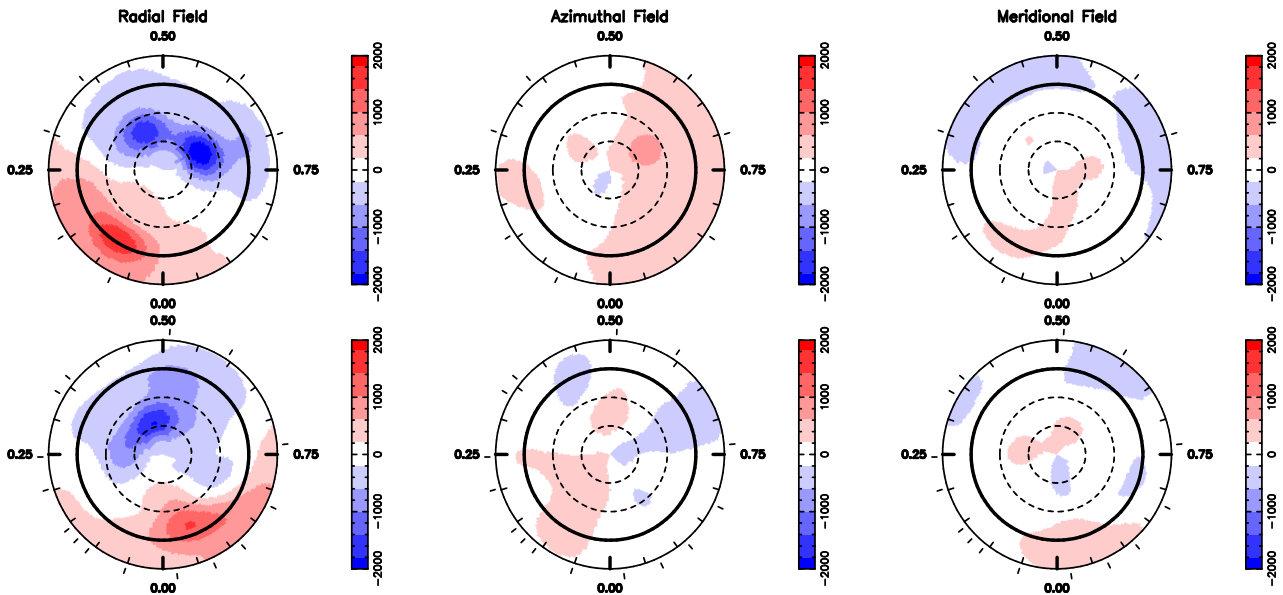

 \center{%
 \includegraphics[scale=0.9]{fig/evlac_V_06db_map.ps}
 \includegraphics[scale=0.9]{fig/evlac_V_07d_map.ps}
 \caption[]{Same as Figure~\ref{fig:zdi_maps_adleo} for EV~Lac, using
data obtained in 2006 (upper row) and 2007 (lower row).}
 \label{fig:zdi_maps_evlac}}
\end{figure*}

\begin{figure}
 \center{%
 \includegraphics[scale=0.4]{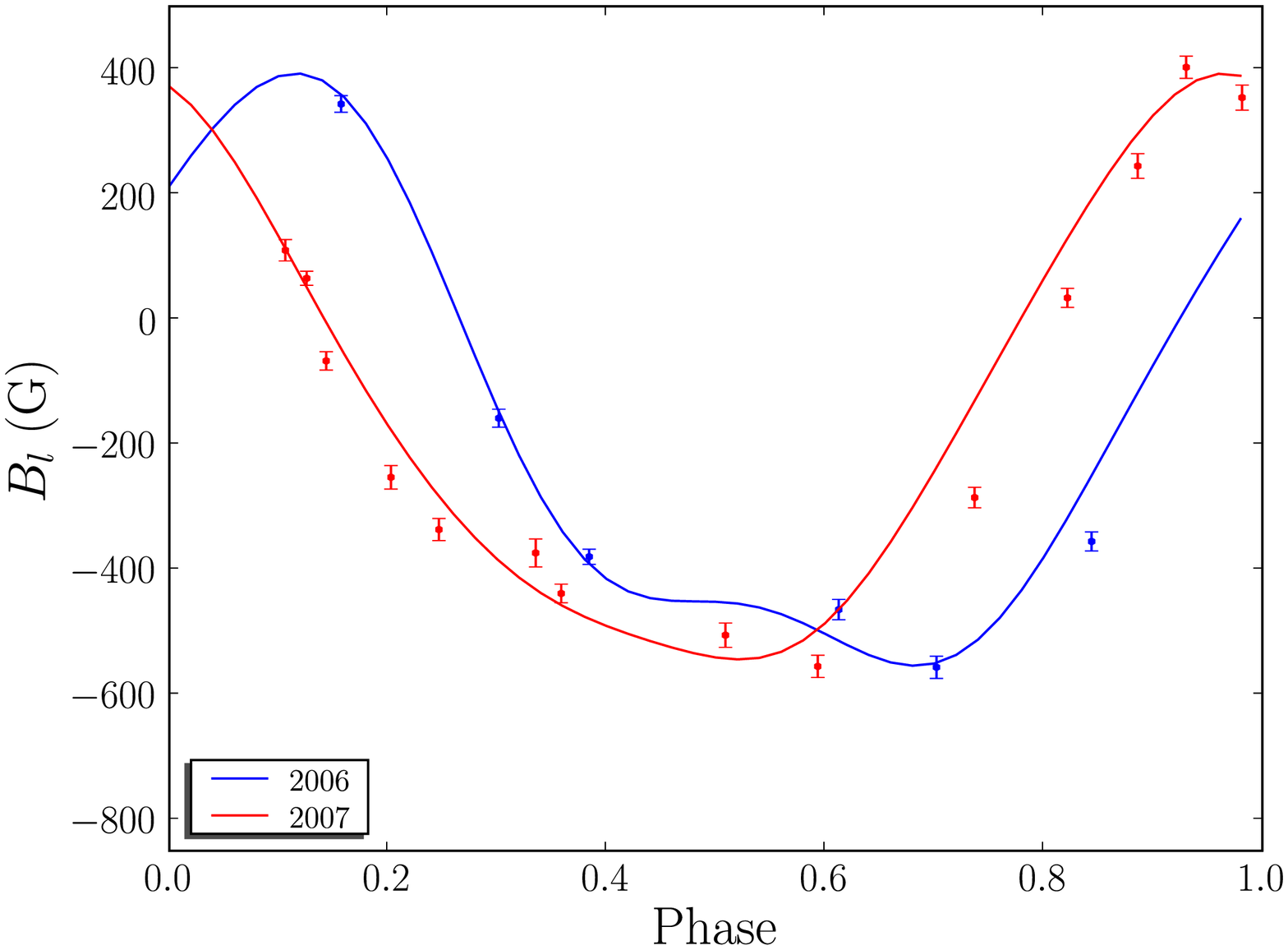}
 \caption[]{Same as Figure~\ref{fig:bl_adleo} for EV~Lac.}
 \label{fig:bl_evlac}}
\end{figure}

EV Lac was observed in August 2006 and July-August 2007, we respectively
obtained 7 and 15 spectra (see Tab.~\ref{tab:log_evlac} and
Fig.~\ref{fig:zdi_spec_evlac}) providing complete though not very dense
phase coverage (See Fig.~\ref{fig:zdi_maps_evlac}) . We detect
strong signatures in all the spectra and modulation is obvious for each
time-series (see Fig.~\ref{fig:bl_evlac}). We measure mean RV of
$0.28~\kms$ and $0.36~\kms$ in 2006 and 2007 respectively, in good
agreement with the value of $0.41\pm0.1~\kms$ reported by
\cite{Nidever02}. The dispersion about these mean RV is equal to
$0.13~\kms$ at both epochs. These RV variations are
smooth and correlate well with longitudinal fields in our 2007 data, but
the correlation is less clear for 2006 (sparser) data. Assuming a
rotation period of $4.378~\d$, determined photometrically by
\cite{Pettersen80}, and considering $\vsini \simeq 3.0~\kms$
\citep{Reiners07} or $\vsini=4.5\pm0.5~\kms$ \citep{Johns96}, we
straightforwardly deduce $\rstar~\sin i \simeq 0.35~\rsun$. As $\rstar
\simeq 0.30~\rsun$, we expect a high inclination angle.

We use the above value for $P_{ZDI}$, $i = 60\degr$, and perform a
spherical harmonics decomposition up to order $\ell=8$. It is then
possible to fit our Stokes $V$ 2007 data set from an initial
$\chisqr=82$ down to $\chisqr = 2.0$ for any velocity $3.0 \leq \vsini
\leq 5.0~\kms$. Neither the fit quality on Stokes $I$ spectra nor the
properties of the reconstructed magnetic topology are significantly
affected by the precise value of \vsini, whereas the filling factors and
the  reconstructed magnetic flux are. The greater the velocity the lower
the filling factors, and the average magnetic flux $B$ ranges from
$0.5~\kG$ at $5.0~\kms$ to $0.6~\kG$ at $3.0~\kms$. Despite the fact
that we achieve a poorer fit for the 2006 data set (from an initial
$\chisqr=125$), $\chisqr = 4.0$ for $\vsini = 5.0~\kms$ and $\chisqr = 4.5$ for
$3.0~\kms$, the same trends are observed. In the rest of the paper we
assume $\vsini = 4.0~\kms$ for EV~Lac.

We recover simple and fairly similar magnetic topologies from both
data sets (see Fig.~\ref{fig:zdi_maps_evlac}). The surface magnetic
field reconstructed from 2007 data is mainly composed of two strong
spots of radial field of opposite polarities where magnetic flux $B$
reaches more than 1.5~\kG. The spots are located at opposite longitudes;
the positive polarity being on the equator and the negative one around
50\degr\ of latitude. The field is far from axisymmetry, as expected
from the polarity reversal observed in Stokes $V$ signature during the
rotation cycle (see Fig.~\ref{fig:zdi_spec_evlac}). The 2006 topology
differs by a rather stronger magnetic flux, maximum flux is above 2~\kG\
with average flux stronger by 0.1~\kG\ than in 2007; the spot of negative
polarity is splitted into two distinct structures; and toroidal field is
not negligible (in particular visible as spot of azimuthal field).

Magnetic energy is concentrated (60\% in 2006, 75\% in 2007) in the
radial dipole modes $\alpha(1;0)$ and $\alpha(1;1)$, no mode of degree
$\ell > 1$ is above the 5\% level, though fitting the data down to
$\chisqr=2.0$ requires taking into account modes up to $\ell=8$.
Toroidal field gathers more than 10\% of the energy in 2006, whereas
they are only marginally reconstructed (2\%) in 2007. Although the
magnetic distribution is clearly not axisymmetric, $m=0$ modes encompass
approximately one third of the magnetic energy at both epochs.

We then try to constrain the surface differential rotation of EV~Lac as
explained in Section~\ref{sec:mod-diffrot}. The \chisqr\ map computed
from 2007 data can be fitted by a paraboloid. We thus infer the
rotation parameters: $\Omeq = 1.4385\pm0.0008~\rpd$ and $\dOm =
1.7\pm0.8~\mrpd$. Our data are thus compatible with solid-body rotation
within 3-$\sigma$. Assuming rigid rotation, we find a clear \chisqr\
minimum for $\Prot=4.37\pm0.01~\d$ (3-$\sigma$ error-bar).

Although the magnetic topologies recovered from 2006 and 2007 are
clearly different, they exhibit common patterns. We merge both data sets
and try to fit them simultaneously with a single magnetic topology.
Assuming solid-body rotation, we find a clear \chisqr\ minimum for
$\Prot=4.3715 \pm 0.0006~\d$ (3-$\sigma$ error-bar). We mention the
formal error bar which may be underestimated since variability can have
biased the rotation period determination. We also find aliases to shifts
of $\sim0.05~\d$, $\Prot=4.3201~\d$ and $\Prot=4.4248~\d$ for the
nearest ones. With $\Delta\chisq$ values of 2522 and 1032 these values
are safely excluded. The periods we find for the 2007 data set alone or
for both data sets are compatible with each other and in good agreement
with the one reported by \cite{Pettersen80} and \cite{Pettersen83}
(4.378 and 4.375~\d) based on photometry.

\section{YZ~CMi = GJ~285 = HIP 37766}
\label{sec:yzcmi}

\begin{figure*}
 \center{%
  \includegraphics[scale=0.45, angle=270]{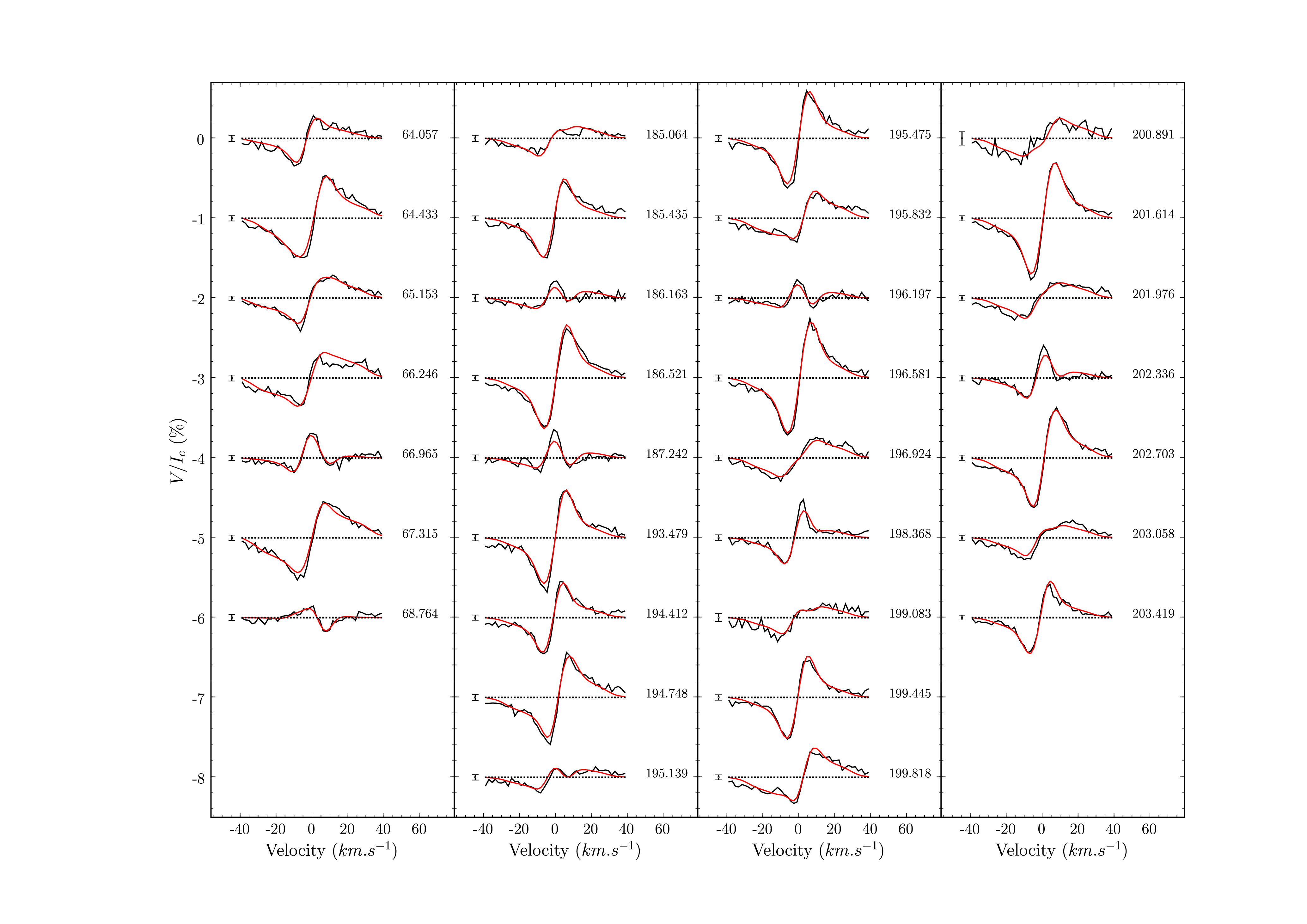}
  \caption[]{Same as Fig.~\ref{fig:zdi_spec_adleo} for YZ~CMi 2006
(column 1) and 2007 (columns 2, 3 and 4) data sets.}
 \label{fig:zdi_spec_yzcmi}}
\end{figure*}

\begin{figure*}
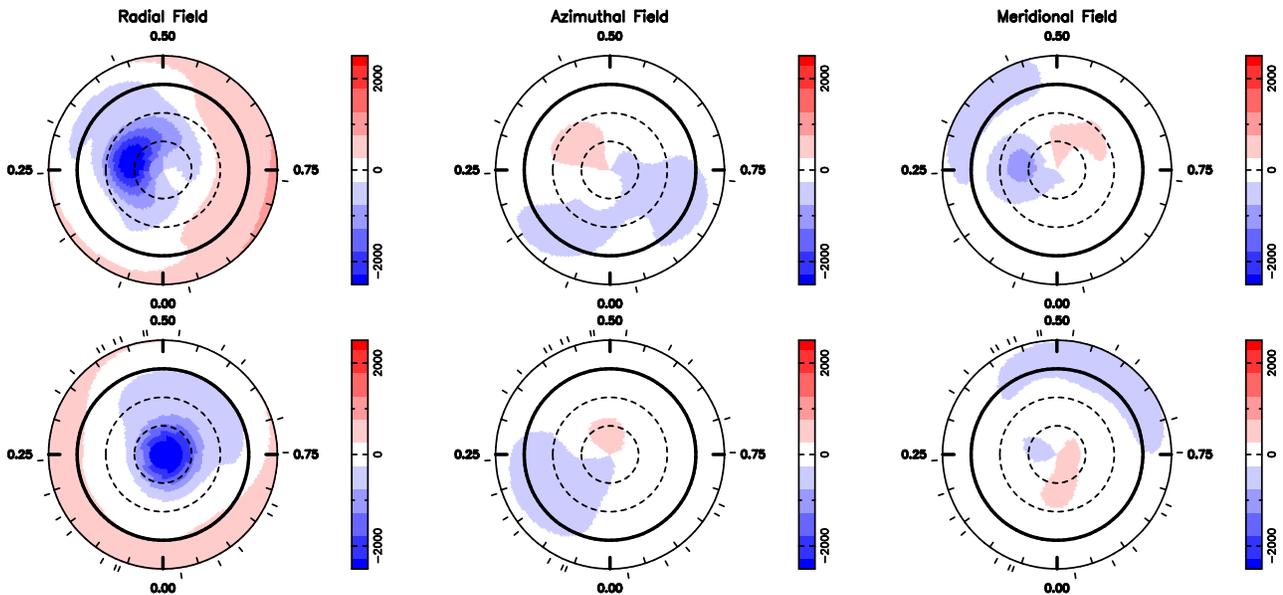

 \center{%
  \includegraphics[scale=0.90]{fig/yzcmi_V_07c_map.ps}
  \includegraphics[scale=0.90]{fig/yzcmi_V_08c_map.ps}
  \caption[]{Same as Figure~\ref{fig:zdi_maps_adleo} for YZ~CMi using
data obtained in 2007 (upper row) and 2008 (lower row).}
 \label{fig:zdi_maps_yzcmi}}
\end{figure*}

\begin{figure}
 \center{%
 \includegraphics[scale=0.4]{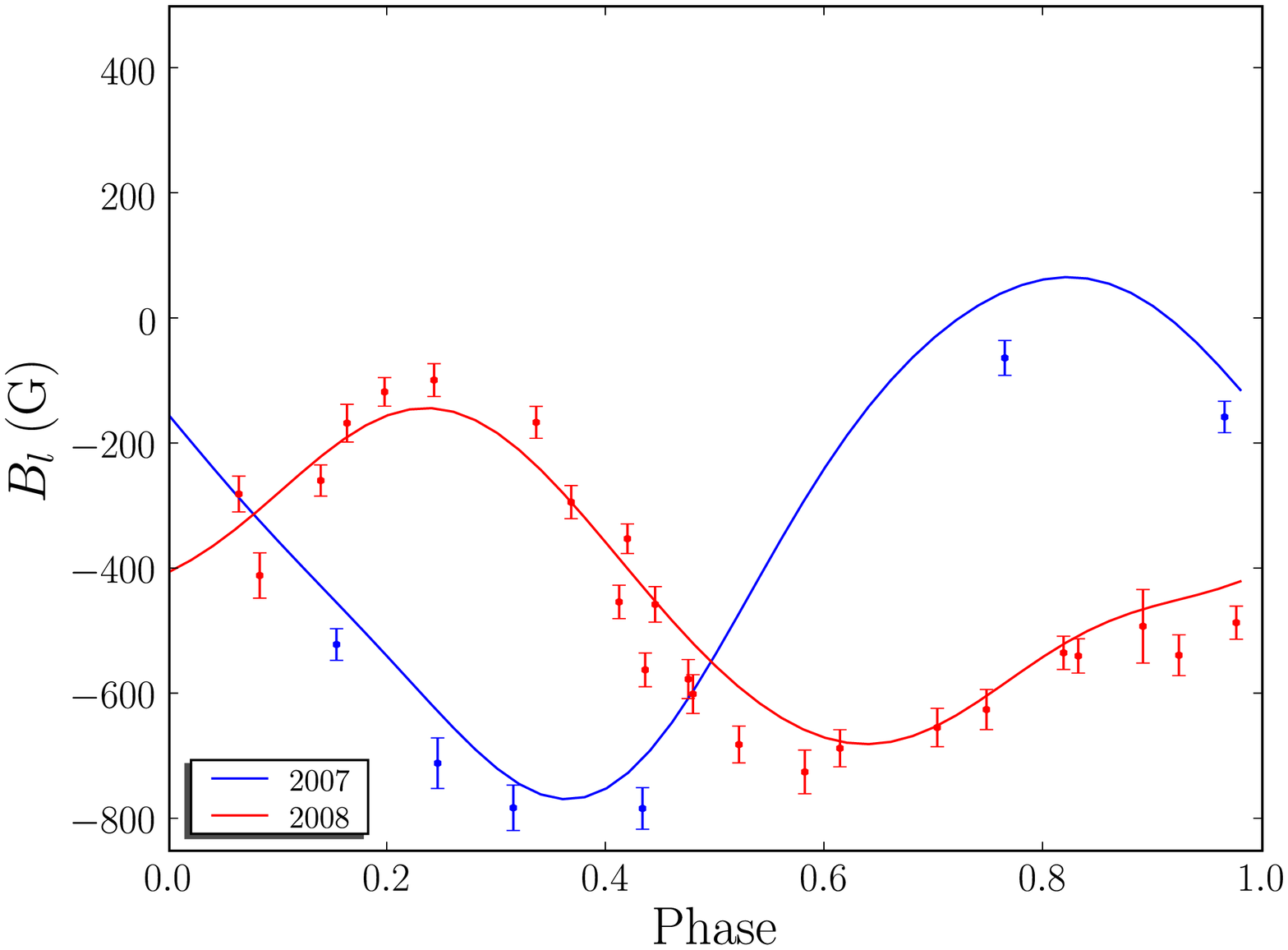}
 \caption[]{Same as Figure~\ref{fig:bl_adleo} for YZ~CMi.}
 \label{fig:bl_yzcmi}}
\end{figure}

We collected 7 spectra of YZ~CMi in January-February 2007 and 25
between December 2007 and February 2008 (see Tab.~\ref{tab:log_yzcmi}
and Fig.~\ref{fig:zdi_spec_yzcmi}). For $P_{ZDI} = 2.77~\d$ \citep[][
photometry]{Pettersen83}, we notice that the 2007 data provide
correct phase coverage for half the rotation cycle only. On the
opposite, the 2008 data provide complete and dense sampling of the
rotational cycle (see Fig.~\ref{fig:zdi_maps_yzcmi}). Rotational
modulation is very clear for both data sets (see 
Fig.~\ref{fig:bl_yzcmi}). We measure mean RV of $26.64~\kms$ and
$26.60~\kms$ in 2007 and 2008 data set, respectively, in good agreement
with $v_r=26.53\pm0.1~\kms$ reported by \cite{Nidever02}. The
corresponding dispersions are $0.13~\kms$ and $0.21~\kms$, the
difference likely reflects the poor phase coverage provided by 2007 data
rather than an intrinsic difference. Although RV varies smoothly with
the rotation phase, we do not find any obvious correlation between
$B_l$ and RV. From the stellar mass (computed from \mj, see
Sec.~\ref{sec:sample}), we infer $\rstar \simeq 0.30~\rsun$. The
above rotation period and $\vsini = 5~\kms$ \citep{Reiners07} implies
$\rsini=0.27~\rsun$ and thus a high inclination angle of the rotational
axis.

We run ZDI on these Stokes $V$ time-series with the aforementioned
values for $P_{ZDI}$ and \vsini, and $i=60\degr$. Both data sets can be
fitted from an initial $\chisqr\simeq38$ down to $\chisqr = 2.0$ using
spherical harmonics decomposition up to order $\ell = 6$. An average
magnetic flux $B \simeq 0.6~\kG$ is recovered for both observation
epochs.

The large-scale topology recovered from 2008 data is quite simple: the
visible pole is covered by a strong spot of negative radial field (field
lines penetrating the photosphere) -- where the magnetic flux reaches up
to 3~\kG\ -- while the other hemisphere is mainly covered by emerging
field lines. Radial, and thus poloidal, field is widely prevailing,
toroidal magnetic energy only stands for 3\% of the whole. The magnetic
field structure also exhibits strong axisymmetry, with about 90\% of the
magnetic energy in $m=0$ modes.

The main difference between 2007 and 2008 maps is that in 2007 this
negative radial field spot is located at a lower latitude. We argue that
this may be partly an artifact due to poor phase coverage. As only one
hemisphere is observed the maximum entropy solution is a magnetic
region facing the observer, rather than a stronger polar spot. We
therefore conclude that non-axisymmetry inferred from 2007 observations
is likely over-estimated.

We try a measurement of differential rotation from our time-series of
Stokes $V$ spectra, as explained in Section~\ref{sec:mod-diffrot}. From
our 2008 data set we obtain a \chisqr\ map featuring a clear paraboloid.
We infer the following rotation parameters: $\Omeq = 2.262\pm0.001\rpd$
and $\dOm = 0.0\pm1.8~\mrpd$. Assuming solid-body rotation, we derive
$\Prot = 2.779\pm0.004~\d$ (3-$\sigma$ error-bar).

We proceed as for AD~Leo to estimate the intrinsic evolution
of the magnetic topology between our 2007 and 2008 observations.
Assuming rigid-body rotation, it is possible to fit the complete data
set down to $\chisqr=3.9$ showing that definite -- though moderate --
variability occurred between the two observation epochs. The rotation
period corresponding to the minimum $\chisqr$ is
$\Prot=2.7758\pm0.0006~\d$ (3-$\sigma$ error-bar). The aliases (shifts
of $\sim0.021~\d$) can be safely excluded ($\Delta\chisq=1450$ and 440
for $\Prot=2.7546~\d$ and $2.7966~\d$, respectively). The periods we
find for the 2008 data set alone or for both data sets are compatible
with each other and with in good agreement the one reported by
\cite{Pettersen83} (2.77~\d) based on photometry.

\section{EQ~Peg~A = GJ~896~A = HIP~116132}
\label{sec:eqpega}

\begin{figure}
 \center{%
  \includegraphics[scale=0.5]{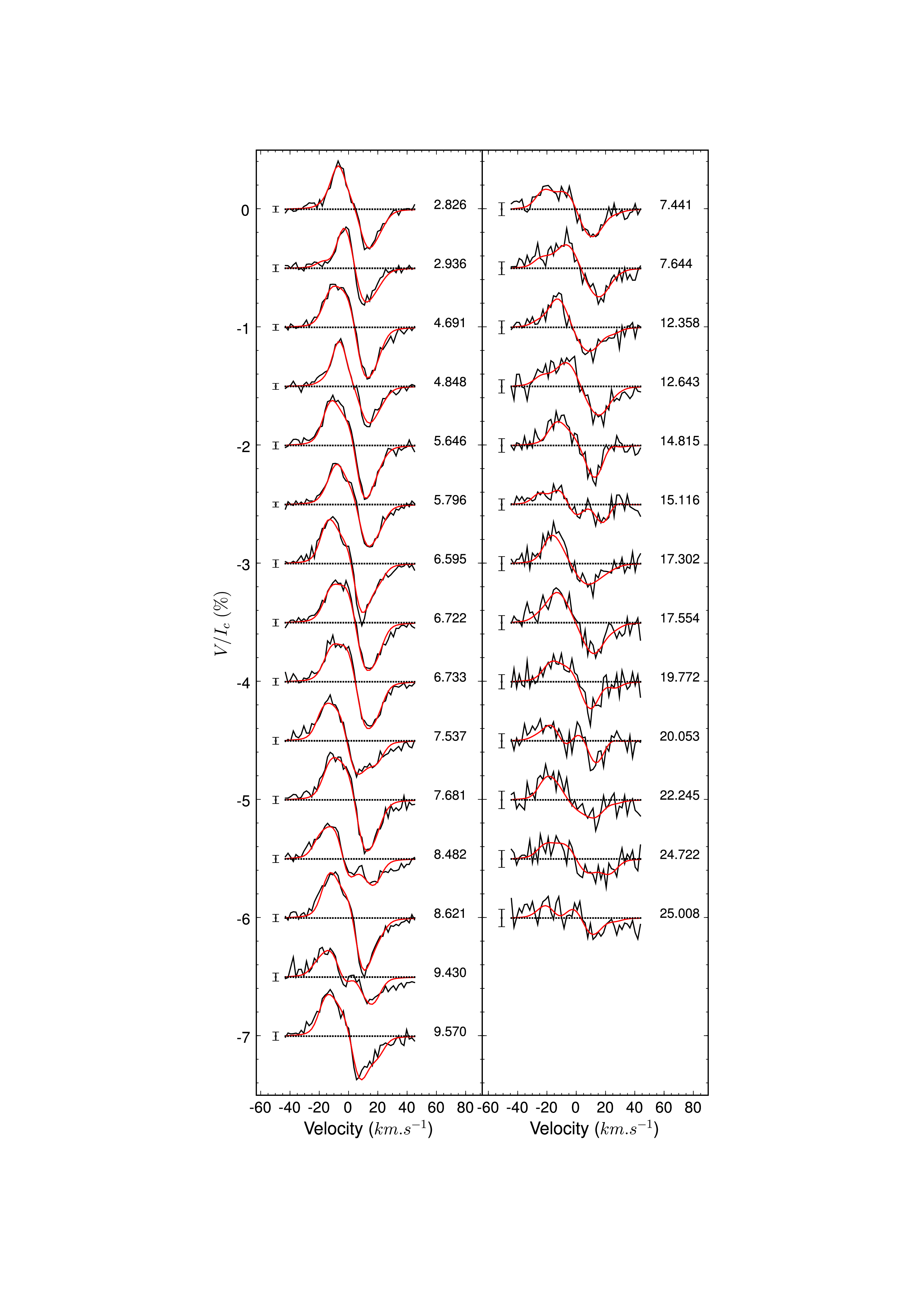}
  \caption[]{Same as Fig.~\ref{fig:zdi_spec_adleo} for EQ~Peg~A
(left-hand column) and EQ~Peg~B (right-hand column) 2006 data sets}
 \label{fig:zdi_spec_eqpegab}}
\end{figure}

\begin{figure*}
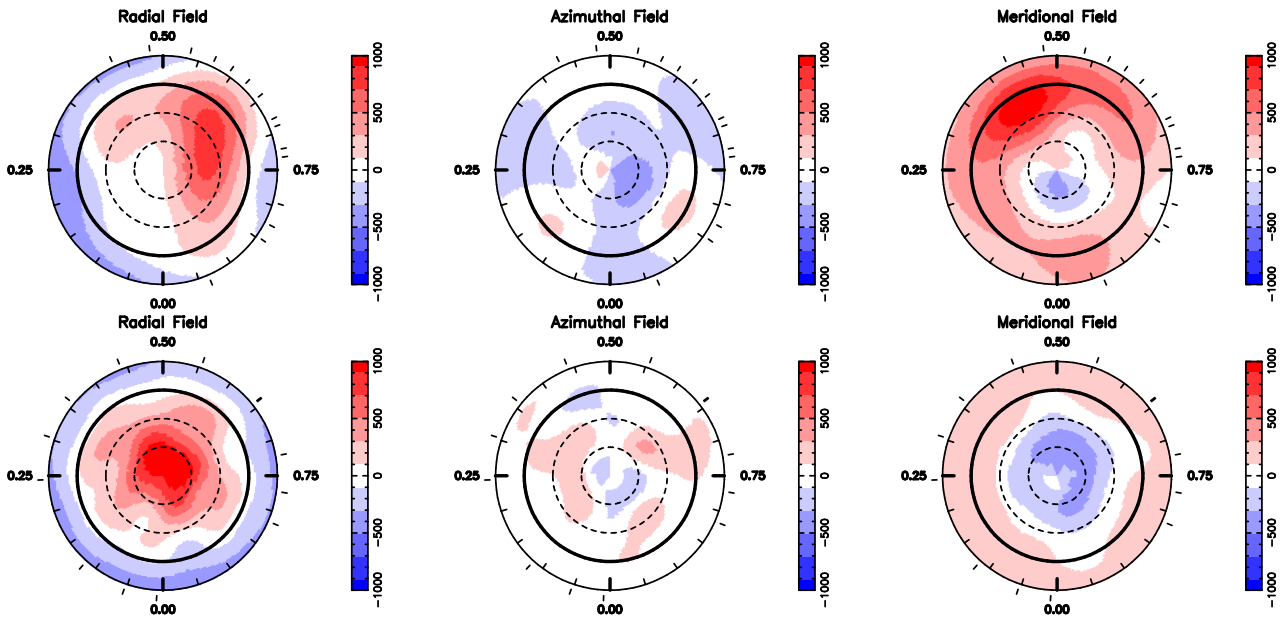

 \center{%
  \includegraphics[scale=0.90]{fig/eqpega_V_06c_map.ps}
  \includegraphics[scale=0.90]{fig/eqpegb_V_06d_map.ps}
  \caption[]{Same as Figure~\ref{fig:zdi_maps_adleo} for EQ~Peg~A (upper
row) and B (lower row) as derived from our 2006 data sets.}
 \label{fig:zdi_maps_eqpegab}}
\end{figure*}

\begin{figure}
 \center{%
 \includegraphics[scale=0.4]{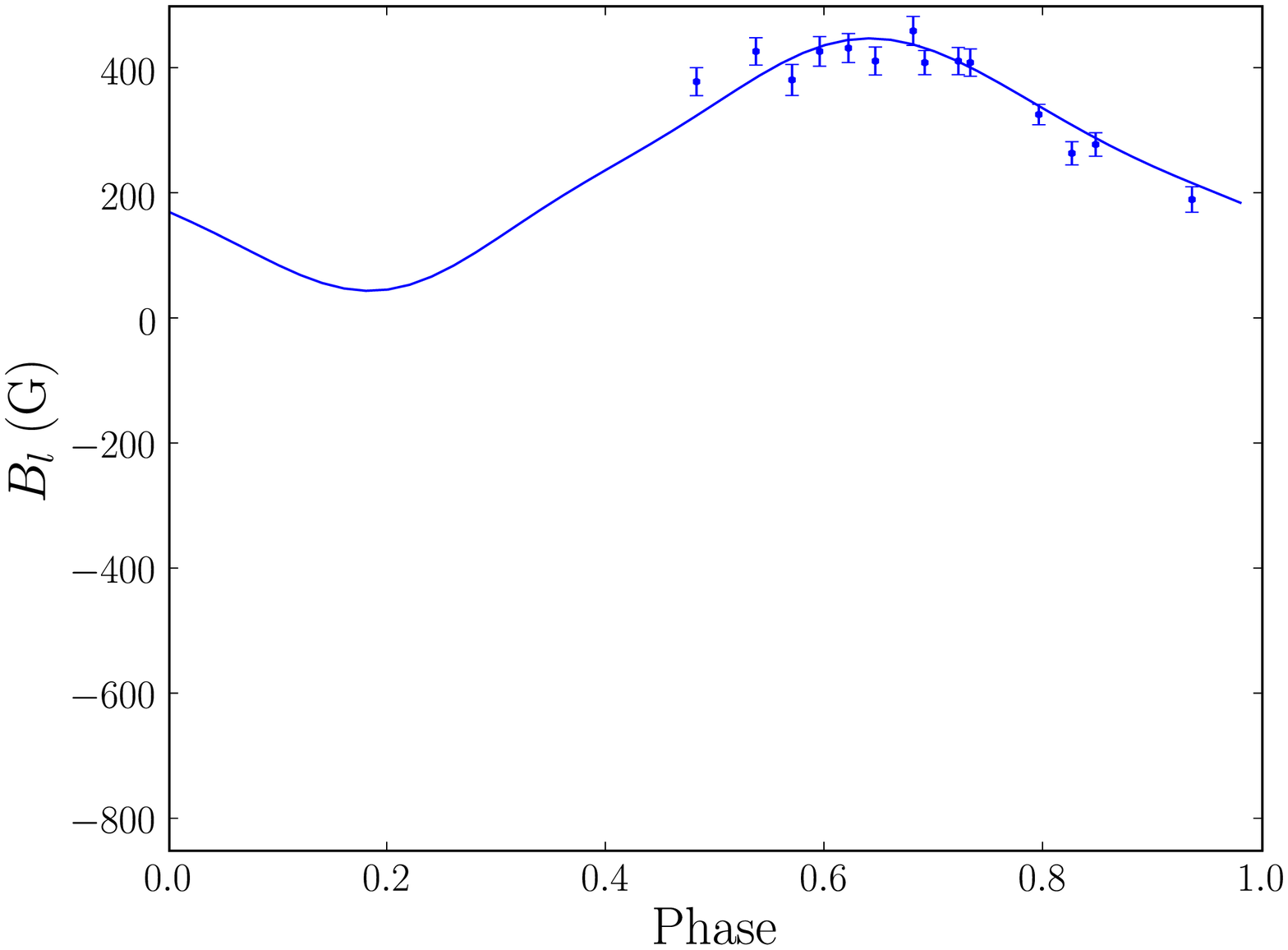}
 \caption[]{Same as Figure~\ref{fig:bl_adleo} for EQ~Peg~A.}
 \label{fig:bl_eqpega}}
\end{figure}

We observed EQ~Peg~A in August 2006 and obtained a set of 15 Stokes $I$
and $V$ spectra (see Tab.~\ref{tab:log_eqpega} and
Fig.~\ref{fig:zdi_spec_eqpegab}), providing observations of only one
hemisphere of the star (see Fig.~\ref{fig:zdi_maps_eqpegab}) considering
$P_{ZDI} = 1.06~\d$. Zeeman signatures are detected in all the
spectra, showing moderate time-modulation (see
Fig.~\ref{fig:bl_eqpega}). We measure a mean RV of $0.31~\kms$ with a
dispersion of $0.18~\kms$. Although RV exhibits smooth
variations along the rotational cycles, we do not find any simple
correlation between RV and $B_l$. We find the best agreement between the
LSD profiles and the model for $\vsini=17.5~\kms$. This implies
$\rstar~\sin i \simeq 0.37~\rsun$, whereas provided $\mstar = 0.39\msun$
we infer $\rstar \simeq 0.35~\rsun$. We thus assume $i=60\degr$ for ZDI
calculations.

Stokes $V$ LSD time-series can be fitted from an initial $\chisqr=44$
down to $\chisqr = 1.5$ using a spherical harmonics decomposition up to
order $\ell=6$ by a field of average magnetic flux $B = 0.5~\kG$.
The recovered magnetic map (see Fig.~\ref{fig:zdi_maps_eqpegab}), though
exhibiting a similar structure of the radial component -- one strong
spot with $B=0.8~\kG$ -- is more complex than those of previous
stars, since we also recover significant azimuthal and meridional fields.

The field is dominated by large-scale modes: dipole modes encompass
70\% of the overall magnetic energy and modes of order $\ell >2$ are all
under the 2\% level. Although poloidal field is greatly dominant, the
toroidal component features 15\% of the overall recovered magnetic
energy. The magnetic topology is clearly not purely axisymmetric but the
$m=0$ modes account for 70\% of the reconstructed magnetic energy.

We use ZDI to measure differential rotation as explained in
Section~\ref{sec:mod-diffrot}. We thus infer $\Omeq = 5.92\pm0.02~\rpd$
and $\dOm = 49\pm43~\mrpd$. This value is compatible with solid body
rotation though the error bar is higher than for EV~Lac and YZ~CMi since
data only span 1 week (rather than about 1 month for previous stars).
Then assuming solid body rotation, we find $\Prot = 1.061\pm
0.004~d$ (3-$\sigma$ error-bar), which is in good agreement with the
period of 1.0664~\d reported by \cite{Norton07}. 

\section{EQ~Peg~B = GJ~896~B}
\label{sec:eqpegb}

\begin{figure}
 \center{%
 \includegraphics[scale=0.4]{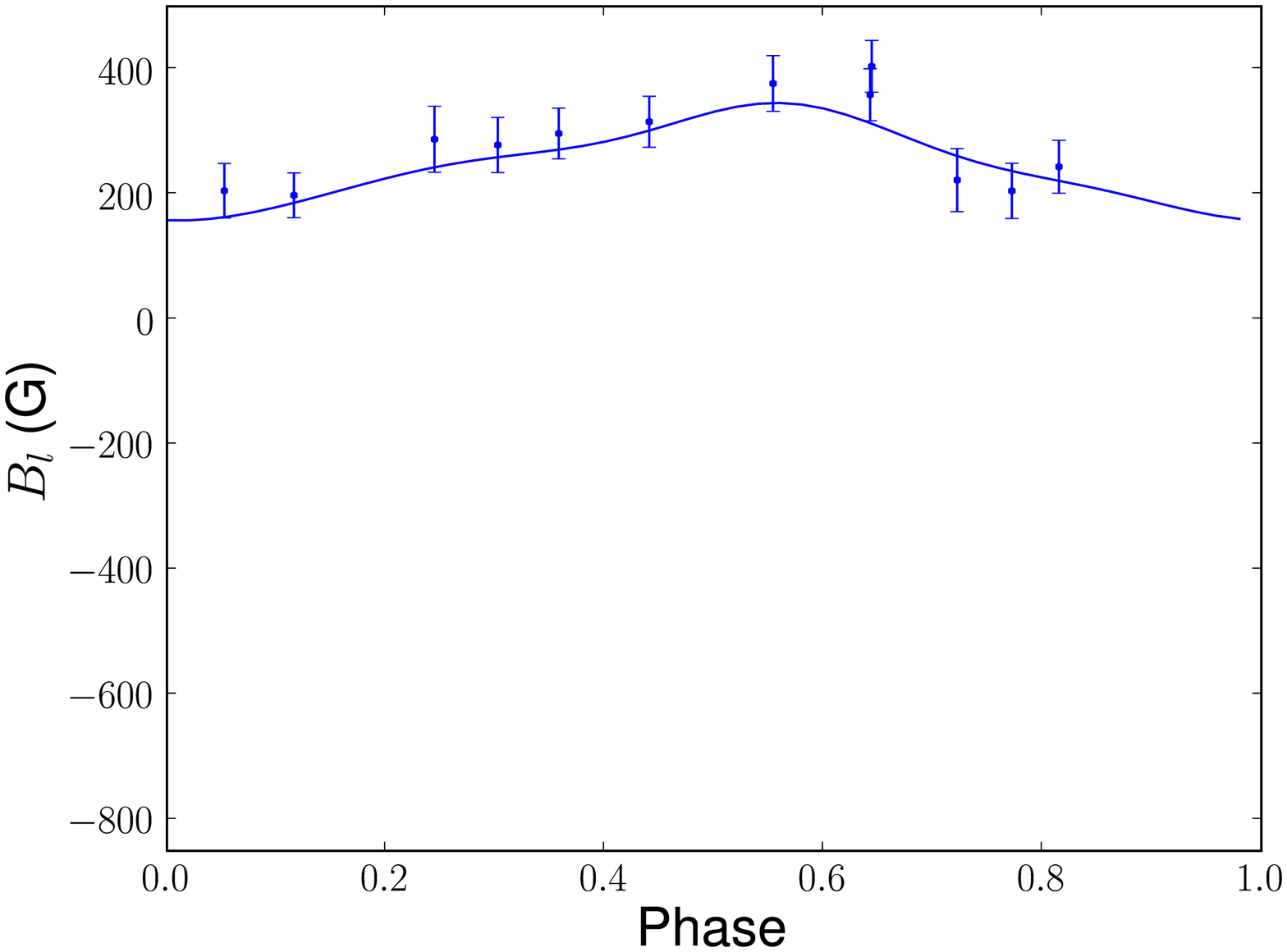}
 \caption[]{Same as Figure~\ref{fig:bl_adleo} for EQ~Peg~B.}
 \label{fig:bl_eqpegb}}
\end{figure}

EQ~Peg~B was observed in August 2006, we obtained a set of 13 Stokes $I$
and $V$ spectra (see Tab.~\ref{tab:log_eqpegb}). Sampling of the
star's surface is almost complete (see Fig.~\ref{fig:zdi_maps_eqpegab})
-- and the derived $P_{ZDI}=0.405~\d$. Stokes $V$ signatures have a
peak-to-peak amplitude above the 1-$\sigma$ noise level in all spectra,
time-modulation is easily detected (see Fig.~\ref{fig:zdi_spec_eqpegab}
and \ref{fig:bl_eqpegb}). We measure a mean RV of $3.34~\kms$ with a
dispersion of $0.16~\kms$. RV is a soft function of
the rotation phase, but we do not find obvious correlation between RV
and $B_l$. We derive a rotational velocity $\vsini = 28.5~\kms$ and thus
$\rsini = 0.23~\rsun$. From the measured J-band absolute magnitude, we
infer $\rstar \simeq 0.25~\rsun$, we will therefore assume $i=60\degr$.

A spherical harmonics decomposition up to order $\ell=8$ allows to fit
the data from an initial $\chisqr=4.6$ down to $\chisqr=1.0$. Using
higher order modes does not result in significant changes. Due to the
high rotational velocity, we find similar results for any value $0<f_V<1$.

The reconstructed magnetic map (see Fig.~\ref{fig:zdi_maps_eqpegab})
exhibits a very simple structure: the hemisphere oriented toward the
observer is mainly covered by positive (emerging) radial fields, in
particular a strong spot ($B = 1.2~\kG$) lies close to the pole; the
other hemisphere is covered by negative radial fields. The meridional
component has the same structure as found for V374~Peg (M08).
Except the (weak) azimuthal component the recovered magnetic topology is
strongly axisymmetric. The average magnetic flux is $B \simeq 0.4~\kG$.

As obvious from Fig.~\ref{fig:zdi_maps_eqpegab} the mode $\alpha(1;0)$
is dominant, it encompasses 75\% of the magnetic energy whereas
no other mode is stronger than 7\%. The field is mostly axisymmetric
with about 90\% of the magnetic energy in $m=0$ modes, and mostly
poloidal ($>95\%$).

Using the method described in Section~\ref{sec:mod-diffrot}, we produce
a map of the \chisqr\ as a function of the rotation parameters \Omeq\
and \dOm\ featuring no clear minimum  in a reasonable range of values.
This may due to a poor constraint on differential rotation since our
data set only span 1 week, and the magnetic topology is mainly composed
of one polar spot. Assuming solid body rotation, we find $\Prot
=0.404\pm0.004~d$ (3-$\sigma$ error-bar).

\section{Discussion and conclusion}
\label{sec:disc}

\begin{table*}
\caption[]{Magnetic quantities derived from our study. For each star,
different observation epochs are presented separately. In columns 2--5
we report quantities from Table~\ref{tab:sample}, respectively the
stellar mass, the rotation period (with an accuracy of 2 digits), the
effective Rossby number and the X-ray to bolometric luminosity ratio.
Columns 6, 7 and 8 mention the Stokes $V$ filling factor, the
reconstructed magnetic energy and the average magnetic flux. Columns
9--13 list the percentage of reconstructed magnetic energy respectively
lying in poloidal, dipole (poloidal and $\ell=1$), quadrupole (poloidal
and $\ell=2$), octupole (poloidal and $\ell=3$) and axisymmetric modes
($m=0$ / $m < \ell/2$).}
 \begin{tabular}{ccccccccccccc}
\hline
Name & Mass & \Prot & $Ro$ & log$R_X$ & $f_V$ & $<B^2>$ & $<B>$ & pol. &
dipole &
quad. &
oct.
&
axisymm. \\
 & (\msun) & (d) & ($10^{-2}$) & & & ($\rm10^5\,G^2$) & (kG) & (\%) &
(\%) & (\%) & (\%) & (\%) \\ 
\hline
EV~Lac (06) & 0.32 & 4.38 & 6.8 & -3.3 & 0.11 & 4.48 & 0.57 & 87 & 60 &
13 & 3 & 33/36\\
\phantom{EV~Lac} (07) & -- & -- & -- & -- & 0.10 & 3.24 & 0.49 & 98 & 75
& 10 & 3 & 28/31\\
YZ~CMi (07) & 0.31 & 2.77 & 4.2 & -3.1 & 0.11 & 5.66 & 0.56 & 92 & 69 &
10 & 5 & 56/61\\
\phantom{YZ~CMi} (08) & -- & -- & -- & -- & 0.11 & 4.75 & 0.55 & 97 & 72
& 11 & 8 & 85/86\\
AD~Leo (07) & 0.42 & 2.24 & 4.7 & -3.2 & 0.14 & 0.61 & 0.19 & 99 & 56 &
12 & 5 & 95/97\\
\phantom{AD~Leo} (08) & -- & -- & -- & -- & 0.14 & 0.61 & 0.18 & 95 & 63
& 9 & 3 & 85/88\\
EQ~Peg~A (06) & 0.39 & 1.06 & 2.0 & -3.0 & 0.11 & 2.73 & 0.48 & 85 & 70 &
6 & 6 & 69/70\\
EQ~Peg~B (06) & 0.25 & 0.40 & 0.5 & -3.3 & na & 2.38 & 0.45 & 97 & 79 & 8
& 5 & 92/94\\
V374~Peg (05) & 0.28 & 0.45 & 0.6 & -3.2 & na & 6.55 & 0.78 & 96 & 72 & 
12 & 7 & 75/76\\
\phantom{V374~Peg} (06) & -- & -- & -- & -- & na & 4.60 & 0.64 & 96 & 70
& 17 & 4 & 76/77\\

\hline
 \label{tab:syn}
 \end{tabular}
\end{table*}

\begin{figure*}
 \center{%
  \includegraphics[scale=0.60]{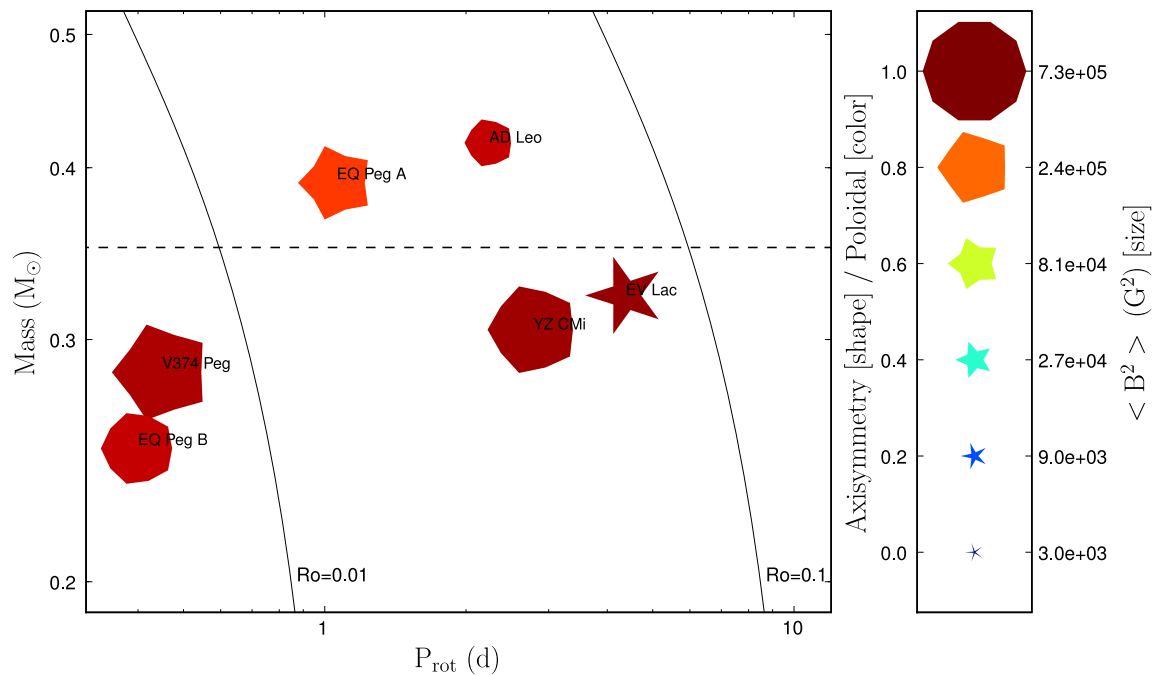}
  \caption[]{Properties of the magnetic topologies of M dwarfs as a
function of rotation period and stellar mass. Larger symbols indicate
larger magnetic fields while symbol shapes depict the different degrees
of axisymmetry of the reconstructed magnetic field (from decagons for
purely axisymmetric fields to sharp stars for purely non axisymmetric fields).
Colours illustrate the field configuration (dark blue for purely toroidal
fields, dark red for purely poloidal fields and intermediate colours for
intermediate configurations). Solid lines represent contours of constant
Rossby number $Ro=0.1$ and $0.01$ respectively corresponding approximately to
the saturation and super-saturation thresholds \citep[e.g.,][]{Pizzolato03}.
The theoretical full-convection limit ($\mstar \simeq0.35\msun$,
\citealt{Chabrier97}) is plotted as a horizontal dashed line.}
\label{fig:plotMP}}
\end{figure*}

Spectropolarimetric observations of a small sample of active M dwarfs
around spectral type M4 were carried out with ESPaDOnS at CFHT and
NARVAL at TBL between 2006 Jan and 2008 Feb. Strong Zeeman
signatures are detected in Stokes $V$ spectra for all the stars of the
sample. Using ZDI, with a Unno-Rachkovsky's model modified by two
filling factors, we can fit our Stokes $V$ time series. It can be seen
on Fig.~\ref{fig:zdi_spec_adleo}, \ref{fig:zdi_spec_evlac},
\ref{fig:zdi_spec_yzcmi} and \ref{fig:zdi_spec_eqpegab} that rotational
modulation is indeed mostly modelled by the imaging code. 

From the resulting magnetic maps, we find that the observed stars
exhibit common magnetic field properties. (a) We recover mainly poloidal
fields, in most stars the observations can be fitted without assuming a
toroidal component. (b) Most of the energy is concentrated in the dipole
modes, i.e. the lowest order modes. (c) The purely axisymmetric
component of the field ($m=0$ modes) is widely dominant except in
EV~Lac. These results confirm the findings of M08, i.e. that magnetic
topologies of fully-convective stars considerably differ from those of
warmer G and K stars which usually host a strong toroidal component in
the form of azimuthal field rings roughly coaxial with the rotation axis
\citep[e.g.,][]{Donati03a}.

Table~\ref{tab:syn} gathers the main properties of the reconstructed
magnetic fields and Figure~\ref{fig:plotMP} presents them in a more
visual way. We can thus suspect some trends: (a) The only
partly-convective stars of the sample, AD~Leo, hosts a magnetic field
with similar properties to the observed fully-convective stars. The only
difference is that compared to fully-convective stars of similar $Ro$,
we recover a significantly lower magnetic flux on AD~Leo, indicating
that the generation of a large-scale magnetic field is more efficient in
fully-convective stars. This will be confirmed in a future paper by
analysing the early M stars of our sample. (b) We do not observe a
growth of the reconstructed large-scale magnetic flux with decreasing
Rossby number, thus suggesting that dynamo is already saturated for
fully-convective stars having rotation periods lower than 5~\d, in
agreement with \cite{Pizzolato03} and \cite{Kiraga07}. Further
confirmation
from stars with $\Prot \gtrsim10~\d$ is needed. This is supported
by the high X-ray fluxes we report, all lying in the saturated part of the
rotation-activity relation with log$R_X\simeq-3$ \citep[e.g.,][]{James00}.
AD~Leo also exhibits a saturated X-ray luminosity despite a significantly weaker
reconstructed magnetic field, indicating that the coronal heating is not
directly driven by the large-scale magnetic field. (c) The only star showing
strong departure from axisymmetry is EV~Lac, i.e. the slowest rotator
(though lying in the saturated regime with $Ro=0.07$). Further
investigation is needed to check if this a general result for
fully-convective stars having $\Prot \gtrsim4~\d$.

The large-scale magnetic fluxes we report here range from 0.2 to 0.8~\kG.
For AD~Leo, EV~Lac and YZ~CMi, previous measurements from Zeeman
broadening of atomic or molecular unpolarised line profiles report
significantly higher overall magnetic fluxes (several \kG)
\citep[e.g.,][]{Saar85, Johns96, Reiners07}. We therefore conclude that
a significant part of the magnetic energy lies in small-scale fields.
Even for the fast rotators EQ~Peg~A and B and V374~Peg for which ZDI is
sensitive to scales corresponding to spherical harmonics up to order
$\ell=\,$12, 20 and 25 (cf. M08), respectively, we reconstruct a large
majority of the magnetic energy in modes of order $\ell\leq3$. This
suggests that the magnetic features we miss with ZDI lie at scales
corresponding to $\ell > 25$ in the reconstructed magnetic fields of mid-M
dwarfs.

Three stars of the sample have been observed at two different epochs
separated by about 1~yr. AD~Leo, EV~Lac, and YZ~CMi exhibit only faint
variations of their magnetic topology during this time gap, the overall
magnetic configuration remained stable similarly to the behaviour of
V374~Peg (cf. M08). This is at odds with what is observed in more
massive active stars, whose magnetic fields reportedly evolve
significantly on time-scales of only a few months
\citep[e.g.,][]{Donati03a}.

For three stars of our sample we are able to measure differential
rotation and find that our data are compatible with solid-body rotation.
In addition, for EV~Lac and YZ~CMi we infer that differential
rotation is at most of the order of a few \mrpd\ i.e. significantly
weaker than in the Sun and apparently lower than in V374~Peg (cf.~M08).
This is further confirmed by the fact that the rotation periods we find 
are in good agreement with photometric periods previously published in 
the literature (whenever reliable).  

This result is consistent with the conclusions of the latest numerical 
dynamo simulations in fully convective dwarfs with $Ro\simeq0.01$ 
\citep{Browning08} showing that 
(i) strong magnetic fields are efficiently produced throughout the 
whole star (with the magnetic energy being roughly equal to the 
convective kinetic energy as expected from strongly helical flows, 
i.e., with small $Ro$)  
and that (ii) these magnetic fields successfully manage to quench 
differential rotation to less than a tenth of the solar shear (as 
a result of Maxwell stresses opposing the equatorward transport of 
angular momentum due to Reynolds stresses).  
However, these simulations predict that dynamo topologies of fully 
convective dwarfs should be mostly toroidal, in contradiction with 
our observations showing strongly poloidal fields in all stars of 
the sample;  the origin of this discrepancy is not clear yet.  

%% {\bf Indeed, it is well known that Maxwell
%% stresses oppose to the equatorward transport of angular momentum
%% by Reynolds' ones. Without magnetic fields, DNS usually lead to a
%% cylinder-like rotation profile inside the star (Taylor-Proudman's
%% theorem) and a fast rotating equator compared to poles at the surface
%% \citep[e.g.,][]{Brun04}. On the contrary, the building of magnetic
%% fields by dynamo leads to Maxwell stresses that oppose to this
%% scenario and a final solid-body rotation can even be achieved for
%% a sufficient field strength, that is, the cylindrical rotation is
%% quenched by large amplitude magnetic fields (see the `Cm'-labelled
%% MHD simulation in Browning 2008). As dynamo theories and simulations
%% predict that equipartition-strength fields are only reachable for
%% weak Rossby numbers (i.e. strongly helical flows), it may explain
%% why solid-body rotations are observed in our M-stars sample. However,
%% a discrepancy remains between our observations and these recent
%% simulations of M-dwarfs, as the strong toroidal fields obtained in
%% the DNS are not observed.}

Our study of Stokes $I$ and $V$ time-series allows to measure both the
rotational period (\Prot) and the projected equatorial velocity (\vsini)
of the sample, from which we can straightforwardly deduce the \rsini.
\Prot\ is well constrained by our data sets (see the error-bars in
Tab.~\ref{tab:sample}), therefore the incertitude on \rsini\ essentially
comes from the determination of \vsini\ ($\sigma\simeq1~\kms$). This
leads to an important incertitude on the \rsini\ deduced for slowly
rotating stars. As explained in M08, for V374~Peg we find a \rsini\
significantly greater than the predicted radius . Here (except for
AD~Leo which is seen nearly pole-on) we find $\rsini \simeq \rstar$ (cf.
Tab.~\ref{tab:sample}), suggesting radii larger than the predicted ones.
This is consistent with the findings of \cite{Ribas06} on eclipsing
binaries, further confirmed on a sample of single late-K and M dwarfs by
\cite{Morales08}, that active low-mass stars exhibit significantly
larger radii and cooler \teff\ than inactive stars of similar masses.
\cite{Chabrier07} proposed in a phenomenological approach that a strong
magnetic field may inhibit convection and produce the observed trends.
This back-reaction of the magnetic field on the star's internal
structure may be associated with the dynamo saturation observed in our
sample (see above), and with the frozen differential rotation predicted
by \cite{Browning08} when the magnetic energy reaches equipartition
(with respect to the kinetic energy).

We also detect significant RV variations in our sample (with
peak-to-peak amplitude of up to 700~\ms). We observe the largest RV variations
on the star having the strongest large-scale magnetic field (YZ~CMi). This
suggests that although the relation between magnetic field measurements and RV
is not yet clear, these smooth fluctuations in RV are due to the magnetic field
and the associated activity phenomena. Therefore, if we can predict the RV
jitter due to a given magnetic configuration, spectropolarimetry may help in
refining RV measurements of active stars, thus allowing to detect planets
orbiting around M dwarfs.

The study presented through this paper aims at exploring the magnetic
field topologies of a small sample of very active mid-M dwarfs, i.e.
stars with masses close the full-convection threshold. Forthcoming
papers will extend this work to both earlier (partly-convective) and
later M dwarfs, in order to provide an insight on the evolution of
magnetic topologies with stellar properties (mainly mass and rotation
period). We thus expect to provide new constraints and better
understanding of dynamo processes in both fully and partly convective
stars.

\section*{ACKNOWLEDGEMENTS} 
We thank the CFHT and TBL staffs for their valuable help throughout
our observing runs. We also acknowledge the referee Gibor Basri for his
fruitful comments.

%% Bibliography
\bibliography{Mstars}

\bibliographystyle{mn2e}

\end{document}